\documentclass[11pt]{article}
\usepackage{draft}
\usepackage{cite}
\usepackage{cancel}
\usepackage{mathrsfs}
\usepackage{enumitem}
\usepackage{xcolor}
\usepackage{caption}  
\usepackage{graphicx} 
\usepackage{float} 
\usepackage{cite}
\usepackage{relsize}
\usepackage{physics}
\usepackage{psfrag}
\usepackage{cancel}
\usepackage{array}
\usepackage{amssymb}
\usepackage{amsmath}
\usepackage[compat=1.1.0]{tikz-feynman}
\usepackage{amsthm}
\usepackage{float}
\usepackage{tikz}
\usetikzlibrary{patterns}
\usetikzlibrary{mindmap,decorations.pathmorphing,backgrounds,positioning,fit}
\usetikzlibrary{decorations.markings}
\usepackage{tikz,lipsum,lmodern}
\usepackage[most]{tcolorbox}
\usepackage{hyperref}
\usepackage{xcolor}
\usepackage{tikz}
\usetikzlibrary{decorations.pathmorphing,patterns}
\usepackage{amsmath}
\usepackage{amssymb}
\usepackage{float}
\usepackage{fixmath}
\usepackage{physics}
\usepackage{slashed}
\usepackage{graphicx}
\usepackage{mathrsfs}
\usepackage{amsbsy}
\usepackage{subfig}
\usepackage{multirow}
\usepackage{hyperref}
\usepackage{bbm}
\definecolor{LightGray}{rgb}{0.8, 0.8, 0.8}

\definecolor{twilightlavender}{rgb}{0.54, 0.29, 0.42}
\definecolor{richmaroon}{rgb}{0.69, 0.19, 0.38}
\definecolor{forestgreen(web)}{rgb}{0.13, 0.55, 0.13}
\definecolor{lava}{rgb}{0.81, 0.06, 0.13}
\hypersetup{
	breaklinks,
	colorlinks,
	citecolor=forestgreen(web),
	filecolor=richmaroon,
	linkcolor=lava,
	urlcolor=twilightlavender
}

\usepackage{cleveref}

\crefformat{section}{\S#2#1#3} 
\crefformat{subsection}{\S#2#1#3}
\crefformat{subsubsection}{\S#2#1#3}

\usepackage{verbatim}
 
\usepackage{color}

\title{From closed to open strings: the tensionless route in Kalb--Ramond background and noncommutativity}

\affiliation[a]{Yau Mathematical Sciences Center (YMSC), Tsinghua University, Beijing 100084, China}
\affiliation[b]{Harish-Chandra Research Institute, A CI of Homi Bhabha National Institute,
Chhatnag Road, Jhunsi, Prayagraj (Allahabad), Uttar Pradesh 211019, India}
\usepackage{orcidlink}
\author[a,\orcidlink{0000-0002-4535-3198}]{Sarthak Duary}\emailAdd{sarthakduary@tsinghua.edu.cn}
\author[b,\orcidlink{0009-0008-2963-2497}]{and Sourav Maji}\emailAdd{souravmaji@hri.res.in}

\abstract{
We study tensionless bosonic strings propagating in the presence of a constant Kalb--Ramond background and show how closed strings undergo a transition into open strings. Working in the intrinsically tensionless theory, we show that the Carrollian limit of the closed-string worldsheet induces a universal gluing of `left'- and `right'-moving oscillators, which is deformed in the presence of the background $B$-field. From the action we derive the mixed boundary conditions, construct the corresponding gluing matrix, and obtain the induced vacuum as a squeezed boundary state. This gives a first-principles realization of the closed-to-open string transition in the tensionless regime. We further extend the analysis to toroidal compactification and show that the worldsheet Bose--Einstein-like condensation mechanism continues to hold in the presence of the $B$-field. Second, we analyze boundary noncommutativity in a unified symplectic framework, both in the tensile theory and in the tensionless regime. In the tensile strings, we show that the boundary symplectic form reproduces the Seiberg--Witten noncommutative parameter. We then study the tensionless limit of this construction and show that the boundary $B$-field term remains as the unique surviving source of the symplectic structure. We derive the same structure directly in the intrinsically tensionless strings, where the inverse boundary symplectic form defines the noncommutative parameter of the null string.}

\begin{document}
\maketitle

\section{Introduction}

The behavior of string theory at extremely high energies has long been a source of deep curiosity and theoretical challenge. One striking feature of this regime is the divergence of the string partition function once the temperature exceeds a critical value-the so-called Hagedorn temperature~\cite{Atick:1988si}. This divergence signals a breakdown of the conventional perturbative description of strings and suggests the onset of a new phase: the Hagedorn phase. In this phase, the effective degrees of freedom are believed to be fundamentally different from those of ordinary, weakly coupled strings.

Remarkably, in this ultra-high-energy limit, scattering amplitudes among strings become different~\cite{Gross:1987kza, Gross:1987ar}.\footnote{Intricate relationships emerge among these amplitudes (see, e.g.,\cite{Gross:1988ue}), hinting at an underlying, highly organized symmetry structure-far richer than what is visible at low energies. This has led to speculations that higher-spin gauge symmetries, characteristic of Vasiliev-type higher-spin gravity theories~\cite{Vasiliev:2004qz}, may become manifest in this limit. 
} This high-energy regime stands in stark contrast to the familiar low-energy limit of string theory, where strings effectively shrink to point-like objects and the theory reduces to supergravity. The transition between these two extremes is governed by a single intrinsic scale: the string length $\alpha'$, which sets the tension $T \sim 1/\alpha'$. In the point-particle limit ($\alpha' \to 0$), strings behave like classical particles with no internal structure. Conversely, the very high-energy regime corresponds to the opposite limit-$\alpha' \to \infty$-where the string tension vanishes. This is known as the tensionless limit ($T \to 0$). To better understand this limit, it is instructive to draw an analogy with point particles.\footnote{For a relativistic point particle, taking the mass to zero forces it to move at the speed of light along null geodesics; its worldline becomes null (lightlike). Similarly, a string’s tension plays the role of ``energy per unit length.'' In the tensionless limit, the string’s worldsheet-its two-dimensional spacetime trajectory-becomes a null surface in spacetime.} The study of such null strings dates back to Schild \cite{Schild:1976vq}, but a systematic action principle was only developed much later. An important formulation was given in~\cite{Isberg:1993av} (hence the acronym ILST), where the authors constructed an action for tensionless strings from a Hamiltonian framework
\begin{equation}
\begin{split}
    S_{\text{ILST}} = \int_{\text{WS}} d^2\xi \, V^a V^b \, \partial_a X^\mu \partial_b X^\nu \, \eta_{\mu\nu}.
\end{split}
\end{equation}
Here, $V^a$ are vector densities on the worldsheet. They replace the combination $T\sqrt{-\gamma}\,\gamma^{ab}$ that appears in the Polyakov action for tensile strings. In the tensionless limit, this replacement takes the precise form
\begin{equation}
\begin{split}
    T\sqrt{-\gamma} \, \gamma^{ab} \quad \longrightarrow \quad V^a V^b.
\end{split}
\end{equation}
The right-hand side is inherently degenerate: its determinant vanishes identically, reflecting the fact that the worldsheet being a null surface.

The action $S_{\text{ILST}}$ retains worldsheet diffeomorphism invariance, so gauge-fixing is necessary to extract physical content. One may choose an analogue of the conformal gauge adapted to the tensionless case. Specifically, setting
\begin{equation}
    V^a = (1,\, 0)~,
\end{equation}
corresponds to fixing the worldsheet coordinates such that time evolution is generated solely along the $\tau$-direction. In this gauge, the action simplifies to
\begin{equation}
\begin{split}
    S_{\text{ILST}}^{\text{gauge-fixed}} 
    &= \int_{\text{WS}} d^2\xi \, \partial_\tau X^\mu \partial_\tau X_\mu.
\end{split}
\end{equation}
This is precisely the action of a free massless scalar field in a Carrollian setting-often referred to as a Carroll scalar. 


In tensile string theory, the residual symmetry after gauge fixing is the Virasoro algebra: a single copy for open strings and two copies (Vir $\otimes$ Vir) for closed strings. In the tensionless case, closed strings exhibit the 2D Conformal Carroll Algebra (CCA)-equivalent to Bondi-Metzner-Sachs algebra in three dimensions (BMS$_3$) \cite{Barnich:2006av}-while open strings give rise to a novel Boundary Conformal Carrollian Algebra (BCCA), reflecting the intrinsically degenerate, Carrollian nature of tensionless dynamics.

The connection to asymptotic symmetries of flat spacetime and its interpretation as the Carrollian contraction of the Virasoro algebra were realized in~\cite{Bagchi:2013bga}. Carrollian conformal structures have become a vibrant area of research, with numerous works exploring their role in tensionless strings \cite{Bagchi:2015nca, Bagchi:2016yyf, Casali:2016atr, Casali:2017zkz, Bagchi:2017cte, Bagchi:2019cay, Bagchi:2020fpr, Bagchi:2020ats}, see path-integral quantization of tensionless (super) string \cite{Chen:2023esw}, see related works \cite{Chen:2025gaz} (see, e.g., \cite{Bagchi:2025vri} for more discussions). 

\subsection*{Tensionless limit as a worldsheet Carroll limit.}
In the tensionless limit, the string becomes long and floppy. This behavior can be realized by applying a singular scaling transformation to the worldsheet coordinates $(\sigma, \tau)$,
\begin{equation}
\tau \to \epsilon \tau, \quad \sigma \to \sigma, \quad \alpha^{\prime} \to \frac{c^{\prime}}{\epsilon}, \quad \epsilon \to 0.
\label{eq:scaling_limit}
\end{equation}

Intuitively, this rescaling causes the spatial direction $\sigma$ to dominate over time on the worldsheet, reflecting the characteristics of an extended, tensionless string. The transformation in Eq.~\eqref{eq:scaling_limit} corresponds to a Carrollian limit on the worldsheet \cite{Duval:2014lpa}. This is also referred to as the ultra-relativistic (UR) limit on the worldsheet-a limit in which speed of light on the worldsheet vanishes. 
In terms of worldsheet velocities $v$ and the worldsheet speed of light $c$, this corresponds to 
\begin{equation}
\frac{v}{c} = \frac{\sigma}{\tau} \to \infty, 
\label{eq:worldsheet_sp}
\end{equation}
which describes the regime where the speed of light on the worldsheet vanishes. This justifies interpreting the tensionless limit as an UR limit on the worldsheet. The appearance of such a Carrollian structure signals that the tensionless worldsheet geometry becomes null.
This limiting procedure induces a contraction of the symmetries of the tensile worldsheet i.e., two copies of the Virasoro algebra (Virasoro $\otimes$ Virasoro) 
resulting in the two-dimensional Carroll algebra. This algebra is isomorphic to the $\mathrm{BMS}_3$ algebra. Tensionless (or null) string formulations have recently been used to probe the microscopic origin of black hole entropy. In particular, \cite{Bagchi:2022iqb} showed that null strings can be used to compute the microstate degeneracy of BTZ black holes. Also, there is an emerging link between tensionless strings and holography in asymptotically flat spacetimes (AFS), see e.g., \cite{Kervyn:2025wsb}. This proposes that the worldsheet of a tensionless string, whose geometry is inherently Carrollian, may furnish a natural setting for flat-space holography. Specifically, it suggests a profound duality: Carrollian symmetries on the tensionless string worldsheet could be holographically dual to Carrollian field theories residing on the null boundary of AFS. This perspective offers a promising pathway toward a first-principles, string-theoretic realization of flat-space holography-transcending bottom-up constructions.\footnote{Flat-space holography has attracted significant attention in recent years, primarily through two complementary frameworks. One is celestial holography, which seeks to describe quantum gravity in four-dimensional asymptotically flat spacetime via a conjectural 2D conformal field theory on the celestial sphere at null infinity \(\mathscr{I}^\pm\) \cite{Strominger:2017zoo,Raclariu:2021zjz,Pasterski:2021rjz,Pasterski:2021raf,Donnay:2023mrd, Duary:2022onm, Duary:2024cqb}. Its core objects-celestial amplitudes \cite{Pasterski:2016qvg, Pasterski:2017kqt,Pasterski:2017ylz}-are obtained by recasting standard scattering amplitudes from momentum eigenstates into boost eigenstates, thereby revealing conformal symmetry. The alternative approach, Carrollian holography, posits a 3D boundary theory at null infinity \(\mathscr{I}\) governed by Carrollian (i.e., BMS) symmetry, which arises in the ultra-relativistic limit \(c \to 0\) of relativistic CFTs \cite{Leblond}. Recent work \cite{Donnay:2022aba} shows that massless scattering amplitudes can be reformulated as Carrollian amplitudes, expressed entirely in terms of asymptotic data on \(\mathscr{I}\). It has been proposed that Carrollian amplitudes arise naturally as the flat-space limit of AdS amplitudes. For recent developments exploring this line of thought see, e.g.,\cite{Stieberger:2022zyk,Hijano:2019qmi,Li:2021snj,deGioia:2022fcn,deGioia:2023cbd,Bagchi:2023fbj,Duary:2022afn,Duary:2023gqg,Bagchi:2023cen,Duary:2024fii,Alday:2024yyj, Duary:2024kxl}.}

A remarkable development in this context is the observation that, in the tensionless limit, open strings can emerge from closed strings~\cite{Bagchi:2019cay}. In this process, the vacuum of the tensile closed string evolves into an induced vacuum satisfying Neumann boundary conditions, which can be identified with the boundary state of an open string. Physically, this phenomenon can be interpreted as a worldsheet phase transition, driven by a Bose--Einstein-like condensation of string modes. This provides a concrete realization of how new degrees of freedom arise in the high-energy regime and offers a dynamical picture of the Hagedorn transition.\\

The goal of the present work is to extend this picture by incorporating a constant background Kalb--Ramond field $B_{\mu\nu}$. The presence of such a background introduces nontrivial couplings between `left'- and `right'-moving sectors of the worldsheet theory, modifying both the boundary conditions and the structure of the induced vacuum. Understanding this interplay is essential for a complete description of the closed-to-open string transition in realistic string backgrounds.

Furthermore, when strings are compactified on tori, they acquire momentum and winding modes. It has been shown that a constant $B$-field affects the spectrum nontrivially only for compactifications with dimension $d \geq 2$~\cite{Banerjee:2023ekd}, consistent with the behavior in tensile string theory. This motivates a careful analysis of tensionless strings in compact backgrounds with nonzero $B$-field, as such settings are crucial for understanding dualities and moduli spaces in string theory.\\

A second central theme of this work is the emergence of noncommutative geometry. It is well known that open strings propagating in a constant $(g,B)$ background exhibit noncommuting endpoint coordinates,
\begin{equation}
[x^i, x^j] = i\,\Theta^{ij},
\end{equation}
leading to noncommutative gauge theories on D-branes~\cite{Seiberg:1999vs,Sheikh-Jabbari:1999cvv,Ardalan:1999av,Schomerus:1999ug,Chu:1999gi,Abouelsaood:1986gd,Callan:1985ia,Fradkin:1985qd,Chu:1998qz,Jing:2003,Braga:2001ci,Jing:2005nq}. This observation, which dates back to the work of Witten \cite{Witten1986},  
and was developed systematically by Seiberg and Witten \cite{Seiberg:1999vs}, Sheikh-Jabbari, Ardalan, Arfaei and Shirzad \cite{Sheikh-Jabbari:1997qke,Ardalan:1998ks,Sheikh-Jabbari:1998aur,Ardalan:1998ce,Sheikh-Jabbari:1999krr}
opened the way to a profound connection between string theory and noncommutative geometry. The Seiberg--Witten map relates the open-string data $(G_{ij},\Theta^{ij})$ to the closed-string background fields $(g_{ij},B_{ij})$, establishing a deep connection between string theory and noncommutative geometry.

While this structure is often derived using worldsheet operator methods \cite{Seiberg:1999vs}, it admits a more geometric interpretation in terms of the symplectic structure of the theory~\cite{Faddeev:1988tj,BarcelosNeto:1992,Chu:1998qz,Jing:2003,Braga:2001ci}. In this approach, the boundary contribution to the action defines a presymplectic two-form whose inverse determines the Poisson brackets of the boundary coordinates, thereby providing a direct origin for noncommutativity.

In the tensionless limit, however, the worldsheet theory undergoes a qualitative restructuring. The degeneracy of the worldsheet metric eliminates the canonical kinetic term, and the standard two-dimensional conformal field theory description is no longer applicable. In particular, the usual decomposition into independent left- and right-moving sectors, which underlies the Virasoro structure of tensile strings, breaks down in this limit. This necessitates a reformulation of noncommutativity that does not rely on the conventional CFT framework.

In this work, we adopt the symplectic perspective to address this issue. We show that, despite the absence of the canonical contribution, the boundary term induced by the background $B$-field defines a well-posed symplectic structure. It yields a noncommutative algebra for the endpoint coordinates, demonstrating that noncommutativity persists as an intrinsic feature of the tensionless theory, governed entirely by its boundary symplectic geometry.

\subsection*{{Main results: part I.}}

In this paper, we derive the boundary conditions of a constant Kalb--Ramond background field that arise in this background and show that they take a mixed form, interpolating between purely Neumann and oblique conditions depending on the strength of the Kalb--Ramond field. 

Then, we construct the induced vacuum state of the theory, which we call the generalized induced vacuum. This state is characterized by a constant matrix that identifies `left'- and `right'-moving oscillators in a background-dependent way. When the Kalb--Ramond field is switched off, this matrix reduces to the identity, and the induced vacuum becomes the familiar Neumann boundary state of an open string. In this sense, the presence of the antisymmetric background deforms the gluing between the two sectors while preserving its overall structure.

We then show that the induced vacuum can be written explicitly as a squeezed state built from pairs of `left'- and `right'-moving oscillators. This state automatically satisfies all the required gluing conditions and represents the physical ground state of the tensionless string in the background field. The construction reveals that, in the Carrollian limit, a closed string effectively behaves as an open string attached to a space-filling brane whose worldvolume geometry is modified by the background field.

Furthermore, we demonstrate that the transformation from the usual closed string vacuum to the induced vacuum can be described by a Bogoliubov transformation acting on the oscillator modes. This transformation mixes oscillators and encodes the emergence of open-string-like degrees of freedom in the tensionless limit. As the tension parameter approaches zero, the closed string vacuum continuously evolves into the induced vacuum, showing explicitly how open string behavior arises from a closed string theory in this regime.

We also extend the analysis to the case where the tensionless string is compactified on a torus with a constant Kalb--Ramond background.

Taken together, these results provide a coherent picture of how tensionless strings interact with constant antisymmetric backgrounds. The tensionless limit corresponds to a Carrollian contraction of the worldsheet geometry. The analysis clarifies the emergence of open string from closed string and establishes that the key features of the induced vacuum and worldsheet condensation persist even in the presence of Kalb--Ramond field.

\subsection*{{Main results: part II.}}

In the second part of this work, we discuss the symplectic and geometric structure of the theory, showing that the tensionless worldsheet in a constant Kalb--Ramond background naturally acquires a noncommutative phase-space geometry. Our goal here is to understand, from first principles, how the intrinsic Carrollian dynamics of the null worldsheet leads to a boundary symplectic form and how this gives rise to noncommutative coordinates for the effective open-string endpoints.

Starting from the gauge-invariant action obtained in the first part, we analyze its canonical structure and identify the relevant symplectic potential associated with variations of the worldsheet embedding coordinates. When we calculate the boundary contribution to the symplectic potential, we find that the only surviving term originates from the background field and it defines a first-order symplectic form along the boundary of the worldsheet. 
The inverse of this symplectic form defines the Poisson bracket on the boundary coordinates, and it has been observed that the endpoints of the tensionless string do not commute. In other words, the background field induces a natural noncommutative geometry on the space probed by the string endpoints. The strength of the noncommutativity depends on the antisymmetric 2-form $B$-field. This construction provides a precise geometric realization of how noncommutativity arises directly from the intrinsic symplectic structure of a null worldsheet.

We then compare this result with the well-known Seiberg--Witten analysis of noncommutative geometry in the tensile regime. In the large $B$-limit, the noncommutative parameter becomes $\mathbf{\Theta}\sim \mathbf{B}^{-1}$, which is the same as the noncommutative parameter computed for the tensionless string. 

\subsection*{{Plan of the paper.}}
In \cref{reviewbfield}, we review the framework of tensionless strings in the presence of a uniform Kalb--Ramond field.  
In \cref{sec:boundary_conditions}, we analyze the generalized induced vacuum condition under a uniform Kalb--Ramond background. The discussion covers the construction of the gluing matrix, the formulation of the squeezed-state boundary state, the associated gluing condition, the Bogoliubov generator, and the physical interpretation of these results.  

Subsequently, in \cref{torustd}, we study tensionless strings in a background Kalb--Ramond field compactified on a torus $T^{d}$.  
This section focuses on boundary conditions, the gluing matrix, the Bogoliubov generator, and the emergence of Bose--Einstein-like condensation on the worldsheet.  
Together, \cref{sec:boundary_conditions} and \cref{torustd} constitute our \textbf{Main Results: Part I}.  

In \cref{sec:symplectic_tensile}, we present a symplectic derivation of the noncommutative parameter in the tensile string case, emphasizing the role of the symplectic form and the resulting noncommutative structure.  
We then extend the analysis in \cref{sec:symplectic_tensionless}, where the noncommutativity of tensionless strings is investigated.  
These two sections represent our \textbf{Main Results: Part II}.  

There are four appendices. Appendix~\ref{appA} discusses the boundary state formalism briefly. Appendix~\ref{appendixA} summarizes key aspects of the symplectic form, Poisson brackets and Moyal-$\star$ product.  
Appendix~\ref{appendixB} reviews the worldsheet derivation of the noncommutative parameter and Appendix~\ref{appendixC} presents the two-point function in tensionless string theory.

\section{Tensionless strings in the presence of a uniform Kalb--Ramond field}
\label{reviewbfield}
We review the analysis of tensionless strings in the presence of a uniform Kalb--Ramond field from \cite{Banerjee:2024fbi}.
We consider a relativistic string moving in a $D$-dimensional flat spacetime with Lorentzian signature, in the presence of a constant antisymmetric Kalb--Ramond field $B_{\mu\nu}$. The Polyakov-like version can be obtained via an auxiliary Lagrangian approach. We begin with the following action
\begin{equation}
S = T \int_{\text{WS}} d^2\xi \left[ \sqrt{-\det\gamma_{ab}} + B_{\mu\nu} \dot{X}^\mu X'^\nu \right],
\label{eq:action}
\end{equation}
where $T$ is the string tension, $\xi^a = (\tau, \sigma)$ are the worldsheet coordinates, and $\gamma_{ab}$ is the induced metric on the two-dimensional worldsheet, defined by
\begin{equation}
\gamma_{ab} = \partial_a X^\mu \partial_b X^\nu \eta_{\mu\nu}.
\label{eq:induced_metric}
\end{equation}
Here, $X^\mu(\tau,\sigma)$ ($\mu = 0,1,\dots,D-1$) represent the embedding functions of the string in spacetime, and $\eta_{\mu\nu}$ is the flat Minkowski metric. We use the convention
\begin{align*}
\dot{} \equiv \frac{\partial}{\partial \tau}, \quad ' \equiv \frac{\partial}{\partial \sigma}.
\end{align*}
Varying the action with respect to $X^\mu$ yields the equation of motion
\begin{equation}
\dot{\Pi}^\mu + K'^\mu = 0,
\label{eq:eom}
\end{equation}
where $\Pi_\mu$ and $K_\mu$ are the canonical momentum and spatial conjugate current, respectively, derived from the Lagrangian density. They are given explicitly by
\begin{equation}
\begin{aligned}
\Pi_\mu &= T \left( \frac{(\dot{X} \cdot X') X'_\mu - X'^2 \dot{X}_\mu}{\sqrt{(\dot{X} \cdot X')^2 - \dot{X}^2 X'^2}} + B_{\mu\nu} X'^\nu \right), \\
K_\mu &= T \left( \frac{(\dot{X} \cdot X') \dot{X}_\mu - \dot{X}^2 X'_\mu}{\sqrt{(\dot{X} \cdot X')^2 - \dot{X}^2 X'^2}} - B_{\mu\nu} \dot{X}^\nu \right).
\end{aligned}
\label{eq:pi_k}
\end{equation}
The canonical momentum satisfies two primary constraints
\begin{equation}
\Pi_\mu X'^\mu = 0,
\qquad
(\Pi_\mu - T B_{\mu\nu} X'^\nu)^2 + T^2 X'^2 = 0.
\label{eq:constraints}
\end{equation}
These constraints reflect the diffeomorphism invariance and gauge structure of the theory on the worldsheet. Before taking the tensionless limit of the action, it is instructive to analyze the structure of the action given in \eqref{eq:action}. At first glance, in the tensionless limit $T \to 0$, the action appears to vanish due to the presence of the square root term. However, one can transition to the Hamiltonian formalism to obtain a finite form of the Polyakov action even as $T \to 0$. In what follows, we demonstrate how this method applies to the tensionless case.
The canonical Hamiltonian associated with the action \eqref{eq:action} can be derived using the expressions for $\Pi_\mu$ from \eqref{eq:pi_k}, leading to
\begin{equation}
\Pi_\mu \dot{X}^\mu - \mathcal{L} = 0.
\label{eq:hamiltonian_constraint}
\end{equation}
This result can also be understood from the reparameterization invariance of the original action \eqref{eq:action}. Consequently, the total Hamiltonian is constructed as a linear combination of the constraints given in \eqref{eq:constraints}
\begin{equation}
\mathcal{H}_T = \rho\, \Pi_\mu X'^\mu + \frac{\lambda}{2} \left\{ (\Pi_\mu - T B_{\mu\nu} X'^\nu)^2 + T^2 X'^2 \right\}.
\label{eq:total_hamiltonian}
\end{equation}
Here, $\lambda$ and $\rho$ are Lagrange multipliers introduced to enforce the constraints. The interpolating Lagrangian is then defined as
\begin{equation}
\begin{aligned}
\mathcal{L}_I &= \Pi_\mu \dot{X}^\mu - \mathcal{H}_T \\
&= \Pi_\mu \dot{X}^\mu - \rho \Pi_\mu X'^\mu - \frac{\lambda}{2} \left[ \Pi^2 + T^2 X'^2 - 2T B_{\mu\nu} \Pi^\mu X'^\nu + T^2 B_{\mu\nu} B^\mu_\rho X'^\nu X'^\rho \right].
\end{aligned}
\label{eq:interpolating_lagrangian}
\end{equation}
In this formulation, $\lambda$ and $\rho$ are treated as auxiliary scalar fields on the worldsheet. Since $\Pi_\mu$ is an auxiliary field, it can be integrated out from the interpolating Lagrangian. After performing this integration, we arrive at the following effective Lagrangian
\begin{equation}
\mathcal{L}_I = \frac{1}{2\lambda} \left[ \dot{X}^2 - 2\rho (\dot{X} \cdot X') + (\rho^2 - \lambda^2 T^2) X'^2 + 2\lambda T B_{\mu\nu} \dot{X}^\mu X'^\nu \right].
\label{eq:effective_lagrangian}
\end{equation}
We now interpret the coefficients of $\dot{X}^2$, $\dot{X} \cdot X'$, and $X'^2$ as components of a two-dimensional worldsheet metric $g^{\alpha\beta}$, such that
\begin{equation}
g^{\alpha\beta} =
\begin{bmatrix}
1 & -\rho \\
-\rho & \rho^2 - \lambda^2 T^2
\end{bmatrix},
\quad
\sqrt{-g} = \sqrt{-\det g_{\alpha\beta}} = \frac{1}{\lambda T}.
\label{eq:metric_components}
\end{equation}
This leads to the Polyakov-form action
\begin{equation}
S_I = \frac{T}{2} \int_{\text{WS}} d^2\xi \left( \sqrt{-g} g^{ab} \partial_a X^\mu \partial_b X^\nu \eta_{\mu\nu} + \epsilon^{ab} B_{\mu\nu} \partial_a X^\mu \partial_b X^\nu \right).
\label{eq:polyakov_action}
\end{equation}
Now, taking the tensionless limit that is, letting $T \to \epsilon T$ with $\epsilon \to 0$, we must rescale the worldsheet variables to maintain a non-degenerate structure. Specifically, we redefine the metric and $B$-field as
\begin{equation}
T \sqrt{-g} g^{ab} \to V^a V^b
\quad \text{and} \quad
B_{\mu\nu} \to \frac{B_{\mu\nu}}{\epsilon},
\quad \text{where} \quad
V^a = \frac{1}{\sqrt{\lambda}} (1, -\rho).
\label{eq:rescaling}
\end{equation}
This rescaling allows us to extract a well-defined limit of the action despite the vanishing string tension.

Here, $V^a$ represents a timelike vielbein density (vector density) on the worldsheet, which encodes the local geometry of the two-dimensional surface. We now introduce a rescaled antisymmetric field $\mathfrak{B}_{\mu\nu}$ defined by
\begin{equation}
\mathfrak{B}_{\mu\nu} = \frac{1}{\alpha'} B_{\mu\nu}, \quad T = \frac{1}{2\pi\alpha'},
\label{eq:rescaled_B}
\end{equation}
where $\alpha'$ is the string length parameter. The scaling of the $B$-field ensures that $\mathfrak{B}_{\mu\nu}$ remains finite in the tensionless limit. With this redefinition, the action for a tensionless string in the presence of a $B$-field becomes
\begin{equation}
S = \int_{\text{WS}} d^2\xi \left( V^a V^b \partial_a X^\mu \partial_b X^\nu \eta_{\mu\nu} + \frac{1}{2\pi} \epsilon^{ab} \mathfrak{B}_{\mu\nu} \partial_a X^\mu \partial_b X^\nu \right).
\label{eq:tensionless_action}
\end{equation}
This expression will serve as the primary object of study in the remainder of this work. It is important to note that this action generalizes the so-called ILST action \cite{Isberg:1993av}, originally proposed for tensionless strings in flat spacetime, by incorporating a consistent coupling to a constant antisymmetric tensor field.

\section{Generalized induced vacuum condition in the presence of a uniform $B$-field}
\label{sec:boundary_conditions}
We consider the tensionless string action in a constant antisymmetric background
\begin{equation}
S \;=\; \int_{\text{WS}} d^2\xi\;\Big(
V^a V^b\,\partial_a X^\mu\partial_b X^\nu\,\eta_{\mu\nu}
\;+\;\frac{1}{2\pi}\,\epsilon^{a b}\,\mathfrak B_{\mu\nu}\,\partial_a X^\mu\partial_b X^\nu
\Big),
\label{eq:action1}
\end{equation}
where $\xi^a=(\tau,\sigma)$ are worldsheet coordinates, $V^a$ is the vector density of the tensionless formalism, $\eta_{\mu\nu}$ is a flat target-space metric, and $\mathfrak B_{\mu\nu}=-\mathfrak B_{\nu\mu}$ is constant. 
We split the Lagrangian density into
\begin{equation}
\mathcal L_1 \equiv V^a V^b \partial_a X^\mu \partial_b X_\mu, 
\qquad
\mathcal L_B \equiv \frac{1}{2\pi}\epsilon^{a b}\mathfrak B_{\mu\nu}\partial_a X^\mu\partial_b X^\nu.
\end{equation}
\paragraph{Kinetic term.}
Varying $\mathcal L_1$ gives
\begin{equation}
\delta \mathcal L_1 = 2 V^a V^b \partial_a X_\mu \,\partial_b(\delta X^\mu).
\end{equation}
Integrating by parts,
\begin{equation}
\delta S_1 = -\int_{\text{WS}} d^2\xi \, 2\partial_b\!\big(V^a V^b \partial_a X_\mu\big)\,\delta X^\mu
+ 2\!\int_{\partial\Sigma} d\tau \, n_b V^a V^b \partial_a X_\mu\,\delta X^\mu,
\end{equation}
where $n_b$ is the outward-pointing normal one-form along $\partial\Sigma$.

\paragraph{$\mathfrak B$-term.}
Varying $\mathcal L_B$ and using antisymmetry of $\mathfrak B$,
\begin{equation}
\delta \mathcal L_B = \frac{1}{\pi}\,\epsilon^{a b}\mathfrak B_{\mu\nu}\,\partial_a\delta X^\mu\,\partial_b X^\nu.
\end{equation}
Integrating by parts,
\begin{equation}
\delta S_B = -\frac{1}{\pi}\int_{\text{WS}} d^2\xi \,\epsilon^{a b}\mathfrak B_{\mu\nu}\partial_a\partial_b X^\nu\,\delta X^\mu
+ \frac{1}{\pi}\int_{\partial\Sigma} d\tau \, n_a \epsilon^{a b}\mathfrak B_{\mu\nu}\partial_b X^\nu \,\delta X^\mu.
\end{equation}
The bulk term vanishes since $\epsilon^{a b}\partial_a\partial_b X^\nu=0$.

\paragraph{Total variation.}
The total boundary contribution is
\begin{equation}
\delta S\big|_{\partial\Sigma} 
= \int_{\partial\Sigma} d\tau \,
\Big[\,2 n_b V^a V^b \partial_a X_\mu
+ \tfrac{1}{\pi} n_a \epsilon^{a b} \mathfrak B_{\mu\nu}\partial_b X^\nu \Big]\delta X^\mu.
\label{eq:boundary-variation}
\end{equation}

\noindent
1. \textbf{Kinetic term.} With $V^a = (1,0)$ and $n_a = (1,0)$ (fixed-$\tau$ boundary), we compute
\begin{equation}
n_b V^b = 1.
\end{equation}
Thus,
\begin{equation}
2 n_b V^a V^b \partial_a X_\mu
= 2\,\partial_\tau X_\mu.
\end{equation}

\noindent
2. \textbf{$\mathfrak{B}$-term.} Using coordinates $(\tau,\sigma)$ with $\epsilon^{\tau\sigma} = +1$, we have
\begin{equation}
n_a \epsilon^{a b} \partial_b
= \partial_\sigma.
\end{equation}
Hence, the second term becomes
\begin{equation}
\tfrac{1}{\pi}\,\mathfrak{B}_{\mu\nu}\,\partial_\sigma X^\nu.
\end{equation}

Putting the kinetic and $\mathfrak B$-field contributions together, the vanishing of the boundary variation,
$$
\delta S\big|_{\partial\Sigma}=0,
$$
implies the boundary condition
\begin{equation}
\partial_\tau X_\mu
+
\frac{1}{2\pi}\,
\mathfrak B_{\mu\nu}\,
\partial_\sigma X^\nu
\bigg|_{\partial\Sigma}
=
0.
\label{eq:boundary_condition_X}
\end{equation}

\paragraph{Boundary state formalism and the gluing condition.}

In standard tensile string theory, boundary conditions imposed on the worldsheet fields can be reinterpreted as operator constraints on boundary states in the closed-string channel. This follows from worldsheet duality,
$$
(\sigma,\tau)_{\text{open}}
\longleftrightarrow
(\tau,\sigma)_{\text{closed}},
$$
under which open-string boundary conditions become conditions imposed at fixed worldsheet time in the closed-string description. For the free bosonic string, Neumann and Dirichlet boundary conditions translate into gluing conditions relating oscillator modes acting on the boundary state \cite{Blumenhagen:2009zz,Blumenhagen:2013fgp}. 

Although the tensionless theory does not possess the conventional left-right decomposition of a two-dimensional conformal field theory, an analogous boundary-state structure still survives in the Carrollian framework \cite{Bagchi:2020fpr, Bagchi:2020ats}. In the present case, the boundary condition \eqref{eq:boundary_condition_X} plays the role of a generalized gluing condition among the tensionless oscillators, thereby defining the induced vacuum state\footnote{For tensionless strings there exist three different vacuum sectors, namely the oscillator, flipped and induced vacua \cite{Bagchi:2020fpr}. In the present work we focus only on the induced vacuum sector, since the closed-to-open transition occurs precisely in this sector. As we shall see, the induced vacuum state acquires the structure of a Neumann boundary state.} in the presence of the background $\mathfrak B$-field. Here the boundary is taken at fixed worldsheet time $\tau$, consistent with the closed-string channel interpretation of the boundary-state formalism. A brief review of the standard boundary state formalism is given in Appendix~\ref{appA}.\\

The gluing condition arising from $\delta S\big|_{\partial\Sigma} = 0$ is,
$$
\partial_\tau X_\mu + \tfrac{1}{2\pi}\,\mathfrak{B}_{\mu\nu}\,\partial_\sigma X^\nu \bigg|_{\partial\Sigma} = 0
\,.
$$
 In the gauge choice $ V^a = (1,0) $, the equation of motion (e.o.m.) for the scalar fields, along with the constraints associated with this gauge, yield the following conditions
\begin{equation}
\partial_\tau^2 X^\mu = 0; \quad \partial_\tau X \cdot \partial_\sigma X = 0; \quad \partial_\tau X \cdot \partial_\tau X =0. 
\end{equation}
In shorthand notation, we have 
\begin{equation}
\ddot{X}^\mu = 0, \quad \dot{X} \cdot X' = 0, \quad \dot{X}^2 = 0.
\end{equation}
where we used the convention
\begin{align*}
\dot{} \equiv \frac{\partial}{\partial \tau}, \quad ' \equiv \frac{\partial}{\partial \sigma}.
\end{align*}

These equations describe the dynamics of the string in the tensionless limit.
We now impose closed string boundary conditions on the worldsheet, requiring that the spacetime coordinates satisfy $ X^\mu(\tau, \sigma + 2\pi) = X^\mu(\tau, \sigma) $, analogous to the case of tensile strings. Under these conditions, the e.o.m. can be solved using a mode expansion of the form
\begin{equation}
X^\mu(\sigma, \tau) = x^\mu + \sqrt{2c'} B_0^\mu \tau + \sqrt{2c'} \sum_{n \neq 0} \frac{i}{n} \left( A_n^\mu - i n \tau B_n^\mu \right) e^{-i n \sigma}.
\end{equation}
Here, $c'$ is a constant with dimensions of $[L]^2$, serving as the tensionless analog of the string coupling constant $\alpha'$ that appears in the standard tensile string mode expansion. 
We define the harmonic oscillators
\begin{equation}
C_n^\mu = \frac{1}{2}(A_n^\mu + B_n^\mu), \quad \widetilde{C}_n^\mu = \frac{1}{2}(-A_{-n}^\mu + B_{-n}^\mu).
\end{equation}
These satisfy
\begin{equation}
[C_m^\mu, C_n^\nu] = m \delta_{m+n,0} \eta^{\mu\nu}, \quad [\widetilde{C}_m^\mu, \widetilde{C}_n^\nu] = m \delta_{m+n,0} \eta^{\mu\nu}.
\end{equation}
We lower indices with the spacetime metric $\eta_{\mu\nu}$ (so $\partial_\tau X_\mu = \eta_{\mu\nu} \partial_\tau X^\nu$).
We first compute the derivatives of the mode expansion.
The $\tau$-derivative is
\begin{equation}
\partial_\tau X^\mu = \sqrt{2c'}\, B_0^\mu + \sqrt{2c'} \sum_{n \neq 0} B_n^\mu e^{-i n \sigma},
\end{equation}
and the $\sigma$-derivative is
\begin{equation}
\partial_\sigma X^\mu = \sqrt{2c'} \sum_{n \neq 0} \left( A_n^\mu - i n \tau B_n^\mu \right) e^{-i n \sigma}.
\end{equation}
Now, inserting these into the boundary condition
\begin{equation}
\partial_\tau X_\mu + \frac{1}{2\pi} \mathfrak{B}_{\mu\nu} \partial_\sigma X^\nu \bigg|_{\partial\Sigma} = 0,
\end{equation}
and dividing by the common factor $\sqrt{2c'}$ yields
\begin{equation}
\eta_{\mu\nu} B_0^\nu + \sum_{n \neq 0} \eta_{\mu\nu} B_n^\nu e^{-i n \sigma}
+ \frac{1}{2\pi} \mathfrak{B}_{\mu\nu} \sum_{n \neq 0} \left( A_n^\nu - i n \tau B_n^\nu \right) e^{-i n \sigma} = 0.
\end{equation}
Now, we equate the constant (zero-frequency) piece and each Fourier coefficient of $e^{-i n \sigma}$.
For the zero mode ($n=0$ term)
\begin{equation}
\eta_{\mu\nu} B_0^\nu = 0
\,.
\end{equation}
For each $n \neq 0$, the coefficient of $e^{-i n \sigma}$ must vanish
\begin{equation}
\frac{1}{2\pi} \mathfrak{B}_{\mu\nu} A_n^\nu = - \left( \eta_{\mu\nu} - \frac{i n \tau}{2\pi} \mathfrak{B}_{\mu\nu} \right) B_n^\nu.
\end{equation}
Now, we derive the generalized induced vacuum condition using the boundary condition  
\begin{equation}
\frac{1}{2\pi} \mathfrak{B}_{\mu\nu} A_n^\nu = -\left( \eta_{\mu\nu} - \frac{i n \tau}{2\pi} \mathfrak{B}_{\mu\nu} \right) B_n^\nu,
\label{eq:boundary}
\end{equation}
and the definitions of the new fields
\begin{equation}
C_n^\mu = \frac{1}{2}(A_n^\mu + B_n^\mu), \quad 
\widetilde{C}_{-n}^\mu = \frac{1}{2}(-A_n^\mu + B_n^\mu).
\label{eq:defs}
\end{equation}
Our goal is to rewrite the boundary condition~\eqref{eq:boundary} entirely in terms of $ C_n^\mu $ and $ \widetilde{C}_{-n}^\mu $.
From the definitions~\eqref{eq:defs}, we solve for $ A_n^\mu $ and $ B_n^\mu $. Adding the two equations
\begin{equation}
B_n^\mu = C_n^\mu + \widetilde{C}_{-n}^\mu.
\label{eq:B_in_C}
\end{equation}
Subtracting the two equations
\begin{equation}
A_n^\mu = C_n^\mu - \widetilde{C}_{-n}^\mu.
\label{eq:A_in_C}
\end{equation}

Now, we substitute equations~\eqref{eq:A_in_C} and~\eqref{eq:B_in_C} into the boundary condition~\eqref{eq:boundary} and simplifying, the equation becomes
\begin{equation}\label{eq:main0}
\left( \eta_{\mu\nu} + \frac{1 - i n \tau}{2\pi} \mathfrak{B}_{\mu\nu} \right) C_n^\nu + \left( \eta_{\mu\nu} - \frac{1 + i n \tau}{2\pi} \mathfrak{B}_{\mu\nu} \right) \widetilde{C}_{-n}^\nu = 0.
\end{equation}
We have rewritten the boundary condition entirely in terms of the variables $ C_n^\mu $ and $ \widetilde{C}_{-n}^\mu $. In the quantum theory, the induced vacuum $\ket{I_{\mathfrak{B}}}$ must satisfy this operator equation as a constraint. We have 
\begin{equation} \label{eq:main}
\Bigg[\left( \eta_{\mu\nu} + \frac{1 - i n \tau}{2\pi} \mathfrak{B}_{\mu\nu} \right) C_n^\nu + \left( \eta_{\mu\nu} - \frac{1 + i n \tau}{2\pi} \mathfrak{B}_{\mu\nu} \right) \widetilde{C}_{-n}^\nu \Bigg] \ket{I_{\mathfrak{B}}} = 0.
\end{equation}
This is a pre-limit condition, the physical induced vacuum condition at the strict tensionless limit is discussed below.

\noindent

\subsection*{{Remarks.}}
Setting $\mathfrak{B}_{\mu\nu} = 0$, the equation simplifies to
\begin{equation}
\eta_{\mu\nu} (C_n^\nu + \widetilde{C}_{-n}^\nu) = 0.
\end{equation}
This implies (for each $\mu$)
\begin{equation}
C_n^\mu + \widetilde{C}_{-n}^\mu = 0.
\end{equation}
This is the simplified boundary condition when the $\mathfrak{B}$-field is absent.

The induced vacuum conditions are
\begin{equation}
(C_n^\mu + \widetilde{C}_{-n}^\mu) |I_0\rangle = 0, \quad \forall n.
\label{eq:vacuum_condition}
\end{equation}
This matches with the result in \cite{Bagchi:2019cay}.

We take the tensionless limit (worldsheet Carroll limit) by scaling $\tau \to \epsilon \tau$ and sending $\epsilon \to 0$. We get,
\begin{equation}
\frac{1 \mp i n \tau}{2\pi} \longrightarrow \frac{1}{2\pi}=a \quad(\text{say}).
\end{equation}

Thus, the coefficients of $C_n^\nu$ and $\widetilde{C}_{-n}^\nu$ become identical and mode-independent. The boundary condition simplifies to,
\begin{equation}
\left[ (\eta_{\mu\nu} + a \mathfrak{B}_{\mu\nu}) C_n^\nu + (\eta_{\mu\nu} - a \mathfrak{B}_{\mu\nu}) \widetilde{C}_{-n}^\nu \right] \ket{I_{\mathfrak{B}}} = 0.
\end{equation}

This is the physical induced vacuum condition.


\subsection{Gluing matrix in the presence of a uniform Kalb--Ramond field}
Our goal is to rewrite the equation 
\begin{equation}
\left[ (\eta_{\mu\nu} + a \mathfrak{B}_{\mu\nu}) C_n^\nu + (\eta_{\mu\nu} - a \mathfrak{B}_{\mu\nu}) \widetilde{C}_{-n}^\nu \right] \ket{I_{\mathfrak{B}}} = 0.
\label{eq:original}
\end{equation}
in the form
\begin{equation}
\big(C_n + R\,\widetilde{C}_{-n}\big)\ket{I_{\mathfrak{B}}} = 0 \qquad (n > 0),
\label{eq:target}
\end{equation}
and determine the gluing matrix $R$.
We define the matrices
\begin{align}
M_+ &= \eta + a \mathfrak{B}, \quad \text{with components } (M_+)_{\mu\nu} = \eta_{\mu\nu} + a \mathfrak{B}_{\mu\nu}, \\
M_- &= \eta - a \mathfrak{B}, \quad \text{with components } (M_-)_{\mu\nu} = \eta_{\mu\nu} - a \mathfrak{B}_{\mu\nu}.
\end{align}
Then, Eq.~\eqref{eq:original} becomes
\begin{equation}
(M_+)_{\mu\nu} C_n^\nu \ket{I_{\mathfrak{B}}} + (M_-)_{\mu\nu} \widetilde{C}_{-n}^\nu \ket{I_{\mathfrak{B}}} = 0.
\end{equation}
This is a vector equation in Lorentz index space (one equation for each $\mu$). We can write
\begin{equation}
(M_+)_{\mu\nu} C_n^\nu \ket{I_{\mathfrak{B}}} = - (M_-)_{\mu\nu} \widetilde{C}_{-n}^\nu \ket{I_{\mathfrak{B}}}.
\end{equation}
Assuming $M_+ = \eta + a \mathfrak{B}$ is invertible, we multiply both sides by the inverse matrix $(M_+^{-1})^{\mu\rho}$
\begin{equation}
\delta^\rho_\nu C_n^\nu \ket{I_{\mathfrak{B}}} = - (M_+^{-1})^{\rho\mu} (M_-)_{\mu\nu} \widetilde{C}_{-n}^\nu \ket{I_{\mathfrak{B}}}.
\end{equation}
Thus,
\begin{equation}
C_n^\rho \ket{I_{\mathfrak{B}}} = - \left[ M_+^{-1} M_- \right]^\rho{}_\nu \widetilde{C}_{-n}^\nu \ket{I_{\mathfrak{B}}}.
\end{equation}
In vector form (suppressing Lorentz indices), this reads
\begin{equation}
C_n \ket{I_{\mathfrak{B}}} = - R\, \widetilde{C}_{-n} \ket{I_{\mathfrak{B}}},
\end{equation}
where we define
\begin{equation}
R = M_+^{-1} M_- = (\eta + a \mathfrak{B})^{-1} (\eta - a \mathfrak{B}).
\end{equation}
We therefore obtain the desired form
\begin{equation}
\big(C_n + R\,\widetilde{C}_{-n}\big)\ket{I_{\mathfrak{B}}} = 0 \qquad (n > 0),
\end{equation}
with the gluing matrix $R$ given by
\begin{equation}
R = (\eta + a \mathfrak{B})^{-1} (\eta - a \mathfrak{B}).
\end{equation}
The matrix $R$ acts on the Lorentz indices of the oscillator modes. Its components are
\begin{equation}
R^\mu{}_\nu = \left[ (\eta + a \mathfrak{B})^{-1} (\eta - a \mathfrak{B}) \right]^\mu{}_\nu.
\end{equation}
So, we see that in the tensionless limit, the gluing matrix $R$ becomes independent of the oscillator level $n$ and the worldsheet time $\tau$. This is a crucial simplification: the identification between $C_n$ and $\widetilde{C}_{-n}$ is now universal across all modes.
It implies that the emergent open string sees a constant, mode-independent background geometry.

\subsection{Squeezed-state boundary state and gluing condition}
We derive the oscillator part of the boundary state $\ket{I_{\mathfrak{B}}}$ and show that it satisfies the gluing condition
\begin{equation}
\big(C_n + R\,\widetilde{C}_{-n}\big)\ket{I_{\mathfrak{B}}} = 0 \qquad (n > 0),
\end{equation}
using only the oscillator algebra and standard vacuum conventions.

\subsection*{{Set up and conventions.}}
The oscillators satisfy the commutation relations
\begin{align}
[C_m^\mu,\, C_n^\nu] &= m\,\delta_{m+n,0}\,\eta^{\mu\nu}, \\
[\widetilde{C}_m^\mu,\, \widetilde{C}_n^\nu] &= m\,\delta_{m+n,0}\,\eta^{\mu\nu},
\end{align}
with all the cross-commutators vanishing
\begin{equation}
[C_m^\mu,\, \widetilde{C}_n^\nu] = 0 \quad \forall\, m,n.
\end{equation}
The squeezed-state boundary ket (ignoring normalization and zero modes) is
\begin{equation}
\ket{I_{\mathfrak{B}}} = \mathcal{N} \exp\!\Big( -\sum_{k>0} \frac{1}{k} C_{-k}^\mu R_{\mu}{}^{\nu} \widetilde{C}_{-k\,\nu} \Big) \ket{0}_c.
\end{equation}
We define the exponent operator
\begin{equation}
X \equiv -\sum_{k>0} \frac{1}{k} C_{-k}^\mu R_{\mu}{}^{\nu} \widetilde{C}_{-k\,\nu},
\end{equation}
so that $\ket{I_{\mathfrak{B}}} = \mathcal{N} e^{X} \ket{0}_c$.
Our goal is to prove
\begin{equation}
(C_n^\mu + R^\mu{}_\nu \widetilde{C}_{-n}^\nu) \ket{I_{\mathfrak{B}}} = 0~~, \quad \text{for all } n > 0.
\end{equation}
\subsection*{{Key commutator computation.}}
We compute $[C_n^\alpha, X]$. Only the term with $k = n$ contributes, since $[C_n^\alpha, C_{-k}^\mu] = n \delta_{k,n} \eta^{\alpha\mu}$
\begin{equation}
\begin{split}
[C_n^\alpha, X] &= -\sum_{k>0} \frac{1}{k} [C_n^\alpha,\, C_{-k}^\mu] R_{\mu}{}^{\nu} \widetilde{C}_{-k\,\nu} \\
&= -\frac{1}{n} [C_n^\alpha,\, C_{-n}^\mu] R_{\mu}{}^{\nu} \widetilde{C}_{-n\,\nu} \\
&= -\frac{1}{n} (n \eta^{\alpha\mu}) R_{\mu}{}^{\nu} \widetilde{C}_{-n\,\nu} \\
&= -\eta^{\alpha\mu} R_{\mu}{}^{\nu} \widetilde{C}_{-n\,\nu}.
\end{split}
\end{equation}
In vector notation, this becomes
\begin{equation}
[C_n,\, X] = -R\, \widetilde{C}_{-n},
\end{equation}
where the left-hand side denotes the vector of component commutators, and the right-hand side is the vector with components $-\eta^{\alpha\mu} R_{\mu}{}^{\nu} \widetilde{C}_{-n\,\nu}$.
\subsection*{{Moving $C_n$ through the exponential.}}
We use the Baker-Campbell-Hausdorff (BCH) identity
\begin{equation}
C_n e^{X} = e^{X} \left( C_n + [C_n, X] + \frac{1}{2!} [[C_n, X], X] + \cdots \right).
\end{equation}
Now, we observe
\begin{itemize}
\item $[C_n, X] \propto \widetilde{C}_{-n}$, a creation operator (since $n > 0$),
\item $X$ is a sum of bilinears in creation operators $C_{-k}, \widetilde{C}_{-k}$ for $k > 0$,
\item all such creation operators commute among themselves at different levels: $[\widetilde{C}_{-n}, \widetilde{C}_{-k}] = 0$ for $n \ne k$, and even for $n=k$, $[\widetilde{C}_{-n}, \widetilde{C}_{-n}] = 0$ (no self-commutator),
\item thus $[ [C_n, X], X ] = 0$, and all higher nested commutators vanish.
\end{itemize}
Therefore, the BCH series terminates after the first term:
\begin{equation}
C_n e^{X} = e^{X} \left( C_n + [C_n, X] \right).
\end{equation}
Substituting $[C_n, X] = -R \widetilde{C}_{-n}$, we get
\begin{equation}
C_n e^{X} = e^{X} \left( C_n - R\, \widetilde{C}_{-n} \right).
\end{equation}
\subsection*{{Action on the vacuum.}}
Acting on $\ket{0}_c$, and using $C_n \ket{0}_c = 0$ for $n > 0$
\begin{equation}
\begin{split}
C_n e^{X} \ket{0}_c &= e^{X} \left( C_n - R\, \widetilde{C}_{-n} \right) \ket{0}_c \\
&= e^{X} \left( 0 - R\, \widetilde{C}_{-n} \ket{0}_c \right) \\
&= -e^{X} R\, \widetilde{C}_{-n} \ket{0}_c.
\end{split}
\end{equation}
Therefore,
\begin{equation}
C_n \ket{I_{\mathfrak{B}}} = \mathcal{N} C_n e^{X} \ket{0}_c = -\mathcal{N} e^{X} R\, \widetilde{C}_{-n} \ket{0}_c.
\end{equation}
\subsection*{{Commutation of $R\,\widetilde{C}_{-n}$ with $e^X$.}}
Now, we consider $R\,\widetilde{C}_{-n} \ket{I_{\mathfrak{B}}} = \mathcal{N} R\,\widetilde{C}_{-n} e^{X} \ket{0}_c$.
We claim that $R\,\widetilde{C}_{-n}$ commutes with $X$, because
\begin{itemize}
\item $\widetilde{C}_{-n}$ is a creation operator (negative mode),
\item $X$ is a sum of terms $C_{-k}^\mu \widetilde{C}_{-k\,\nu}$,
\item $[\widetilde{C}_{-n}, C_{-k}] = 0$ since left- and right-movers commute,
\item $[\widetilde{C}_{-n}, \widetilde{C}_{-k}] = 0$ for all $k > 0$, because $-n -k < 0 \ne 0$, so no commutator survives,
\item hence $[\widetilde{C}_{-n}, X] = 0$, and so $[R\,\widetilde{C}_{-n}, X] = 0$.
\end{itemize}
Thus, $R\,\widetilde{C}_{-n}$ commutes with $e^{X}$
\begin{equation}
R\,\widetilde{C}_{-n} e^{X} = e^{X} R\,\widetilde{C}_{-n}.
\end{equation}
Therefore
\begin{equation}
R\,\widetilde{C}_{-n} \ket{I_{\mathfrak{B}}} = \mathcal{N} R\,\widetilde{C}_{-n} e^{X} \ket{0}_c = \mathcal{N} e^{X} R\,\widetilde{C}_{-n} \ket{0}_c.
\end{equation}

\subsection*{{Final result: gluing condition.}}

Now, we compute the full gluing condition
\begin{equation}
\begin{split}
(C_n + R\,\widetilde{C}_{-n}) \ket{I_{\mathfrak{B}}} &= C_n \ket{I_{\mathfrak{B}}} + R\,\widetilde{C}_{-n} \ket{I_{\mathfrak{B}}} \\
&= -\mathcal{N} e^{X} R\,\widetilde{C}_{-n} \ket{0}_c + \mathcal{N} e^{X} R\,\widetilde{C}_{-n} \ket{0}_c \\
&= 0.
\end{split}
\end{equation}
Hence, we conclude
\begin{equation}
\big(C_n + R\,\widetilde{C}_{-n}\big)\ket{I_{\mathfrak{B}}} = 0 \quad \text{for all } n > 0.
\end{equation}
as desired.


\subsection*{{Consistency checks.}}
When the $\mathfrak{B}$-field is absent, the conditions for the induced vacuum are expressed as
\begin{equation}
(C_n^\mu + \widetilde{C}_{-n}^\mu) |I_0\rangle = 0, \quad \forall n.
\end{equation}
This corresponds to the condition of a Neumann boundary state, and its solution is given by
\begin{equation}
|I_0\rangle = \mathcal{N} \prod_{n=1}^{\infty} \exp\left( -\frac{1}{n} C_{-n} \cdot \widetilde{C}_{-n} \right) |0\rangle_c,
\end{equation}
where $ \mathcal{N} $ is a normalization constant.

By definition,
\begin{equation}
R^\mu{}_\nu = \big[(\eta + a \mathfrak{B})^{-1}(\eta - a \mathfrak{B})\big]^\mu{}_\nu.
\end{equation}
Setting $\mathfrak{B} = 0$, we obtain
\begin{equation}
R^\mu{}_\nu\Big|_{\mathfrak{B}=0} = (\eta^{-1}\eta)^\mu{}_\nu = \delta^\mu{}_\nu,
\end{equation}
so $R \to 1$. Therefore, the $\mathfrak{B}$-field boundary condition
\begin{equation}
(C_n^\mu + R^\mu{}_\nu \widetilde{C}_{-n}^\nu) \ket{I_{\mathfrak{B}}} = 0~~,
\end{equation}
reduces to
\begin{equation}
(C_n^\mu + \widetilde{C}_{-n}^\mu) \ket{I_0} = 0,
\end{equation}
which is the Neumann boundary condition.
We define the squeezed operator
\begin{equation}
A = -\sum_{k>0} \frac{1}{k} C_{-k}^\rho R_{\rho}{}^{\sigma} \widetilde{C}_{-k\,\sigma},
\qquad
\ket{I_{\mathfrak{B}}} = \mathcal{N} e^{A} \ket{0}_c.
\end{equation}
We use the BCH identity
\begin{equation}
e^{A} C_n^\mu e^{-A} = C_n^\mu + [A, C_n^\mu] + \frac{1}{2}[A,[A, C_n^\mu]] + \cdots~~.
\end{equation}
Since $A$ is bilinear in negative-frequency oscillators ($-k < 0$) and $C_n^\mu$ has mode $n > 0$, only the first commutator survives (higher-order commutators vanish due to mode mismatch).
Using the oscillator algebra
\begin{equation}
[C_m^\alpha, C_p^\beta] = m\,\eta^{\alpha\beta} \delta_{m,-p},
\end{equation}
we find for $k > 0$
\begin{equation}
[C_{-k}^\rho, C_n^\mu] = -k\,\eta^{\rho\mu} \delta_{k,n}.
\end{equation}
Raising the index using $\eta^{\rho\mu}$, this becomes
\begin{equation}
[A, C_n^\mu] = R^{\mu}{}_{\sigma} \widetilde{C}_{-n}^{\sigma}.
\end{equation}
Thus,
\begin{equation}
e^{A} C_n^\mu e^{-A} = C_n^\mu + R^{\mu}{}_{\sigma} \widetilde{C}_{-n}^{\sigma}.
\end{equation}
Acting on the squeezed state
\begin{equation}
\begin{split}
\big(C_n^\mu + R^{\mu}{}_{\sigma} \widetilde{C}_{-n}^{\sigma}\big) \ket{I_{\mathfrak{B}}}
&= e^{A} C_n^\mu e^{-A} \cdot e^{A} \ket{0}_c \\
&= e^{A} C_n^\mu \ket{0}_c.
\end{split}
\end{equation}
Therefore,
\begin{equation}
\big(C_n^\mu + R^{\mu}{}_{\sigma} \widetilde{C}_{-n}^{\sigma}\big) \ket{I_{\mathfrak{B}}} = 0,
\qquad (n > 0),
\end{equation}
as required.
As $\mathfrak{B} \to 0$, $R \to 1$, so the state becomes
\begin{equation}
\ket{I_{\mathfrak{B}}} \xrightarrow{\mathfrak{B} \to 0} \mathcal{N} \exp\!\left( -\sum_{k>0} \frac{1}{k} C_{-k} \cdot \widetilde{C}_{-k} \right) \ket{0}_c,
\end{equation}
which is precisely the Neumann boundary state. The annihilation proof reduces accordingly: with $R = 1$, we have $[A, C_n^\mu] = \widetilde{C}_{-n}^\mu$, and hence
\begin{equation}
(C_n^\mu + \widetilde{C}_{-n}^\mu) |I_0\rangle = 0.
\end{equation}
We have shown that
\begin{itemize}

\item $R \to 1$ as $\mathfrak{B} \to 0$,
    \item the state $\ket{I_{\mathfrak{B}}}$ reduces to the Neumann boundary state.
\end{itemize}
Hence, the Neumann boundary states are fully consistent in the $\mathfrak{B} \to 0$ limit.

\subsection{Bogoliubov generator}
Now, let us review the analysis of the transition from closed to open string in \cite{Bagchi:2019cay}. The transition from closed to open string can be understood through a Bogoliubov transformation on the worldsheet. Specifically, the relation between the tensionless closed string oscillators $C_n^\mu$ and the tensile closed string oscillators $\alpha_n^\mu$ is given by
\begin{equation}
    \alpha_n^\mu = e^{iG} C_n e^{-iG} = \cosh\theta \, C_n^\mu - \sinh\theta \, \widetilde{C}_{-n}^\mu,
    \label{eq:alpha_C_transform}
\end{equation}
\begin{equation}
    \widetilde{\alpha}_n^\mu = e^{iG} \widetilde{C}_n e^{-iG} = -\sinh\theta \, C_{-n}^\mu + \cosh\theta \, \widetilde{C}_n^\mu,
    \label{eq:alphatilde_C_transform}
\end{equation}
where, 
\begin{equation}
\begin{split}
\cosh\theta &= \frac{1}{2}\left( \sqrt{\epsilon} + \frac{1}{\sqrt{\epsilon}} \right) \\
\sinh\theta &= \frac{1}{2}\left( \sqrt{\epsilon} - \frac{1}{\sqrt{\epsilon}} \right).
\end{split}
\end{equation}
and the generator $G$ is defined as
\begin{equation}
    G = i \sum_{n=1}^{\infty} \theta \left[ C_{-n} \cdot \widetilde{C}_{-n} - C_n \cdot \widetilde{C}_n \right],
    \quad
    \tanh\theta = \frac{\epsilon - 1}{\epsilon + 1}.
    \label{eq:generator_G}
\end{equation}

Because the tensionless closed string oscillators $C_n^\mu$ and $\widetilde{C}_n^\mu$ are linear combinations of both raising and lowering operators of the tensile string, the vacuum state $|0\rangle_c$ annihilated by all mode $C_n^\mu$ and $\widetilde{C}_n^\mu$ operators,
\begin{equation}
\begin{split}
|\text{tensionless closed string vacuum}\rangle \equiv |0\rangle_c : C_n^\mu |0\rangle_c = 0 = \widetilde{C}_n^\mu |0\rangle_c \quad \forall\, n > 0,
\end{split}
\end{equation}
differs from the standard closed string tensile vacuum $|0\rangle_\alpha$, which is defined by the annihilation conditions
\begin{equation}
\begin{split}
|\text{tensile closed string vacuum}\rangle \equiv  |0\rangle_\alpha : \alpha_n^\mu |0\rangle_\alpha = 0 = \widetilde{\alpha}_n^\mu |0\rangle_\alpha \quad \forall\, n > 0.
\end{split}
\end{equation}

\medskip

The tensile closed string vacuum state $|0\rangle_\alpha$ is related to the tensionless closed string vacuum state $|0\rangle_c$ through 
\begin{equation}
    |0\rangle_\alpha = \sqrt{\cosh\theta} \prod_{n=1}^{\infty} \exp\left[ \frac{\tanh\theta}{n} \, C_{-n} \cdot \widetilde{C}_{-n} \right] |0\rangle_c.
    \label{eq:squeezed_vacuum}
\end{equation}

\medskip

To understand how an open string emerges from a closed one, consider the parameter $\epsilon$, which governs the tension scale. When $\epsilon = 1$, we have $\tanh\theta = 0$, and the vacuum reduces to the original closed string vacuum: $|0\rangle_\alpha = |0\rangle_c$. As $\epsilon$ deviates from unity, the vacuum evolves under the action of the Bogoliubov transformation - it becomes increasingly “squeezed,” as shown in Eq.~(\ref{eq:squeezed_vacuum}). In the limit $\epsilon \to 0$, corresponding to the tensionless limit, we find $\tanh\theta \to -1$. The vacuum then becomes,

\begin{equation}
    |0\rangle_\alpha = \mathcal{N} \prod_{n=1}^{\infty} \exp\left[ -\frac{1}{n} C_{-n} \cdot \widetilde{C}_{-n} \right] |0\rangle_c,
    \label{eq:tensionless_vacuum}
\end{equation}

which is precisely the \emph{induced vacuum} $|I_0\rangle$ introduced. This state satisfies Neumann boundary conditions in all spatial directions - the defining property of a space-filling D25-brane in bosonic string theory. Thus, an \emph{open string}, free to move in all dimensions, arises naturally from a \emph{closed string} by taking the \emph{tensionless limit}.\\

To understand how an open string can emerge from a closed string in the tensionless limit, consider the deformation parameter $\epsilon$, which controls the effective string tension via the scaling $\alpha' \sim 1/\epsilon$. The relation between the tensile and tensionless oscillators is encoded in a Bogoliubov transformation parametrized by $\theta(\epsilon)$, where
\[
\cosh\theta = \frac{1}{2}\left( \sqrt{\epsilon} + \frac{1}{\sqrt{\epsilon}} \right), \qquad
\sinh\theta = \frac{1}{2}\left( \sqrt{\epsilon} - \frac{1}{\sqrt{\epsilon}} \right),
\]
so that
\[
\tanh\theta = \frac{\sinh\theta}{\cosh\theta} = \frac{\epsilon - 1}{\epsilon + 1}.
\]

When $\epsilon = 1$ (finite tension), we have $\tanh\theta = 0$, and the vacuum is the standard \textit{tensile closed string vacuum}
\begin{equation}
\begin{split}
|\text{tensile closed string vacuum}\rangle \equiv |0\rangle_\alpha : \quad 
\alpha_n^\mu |0\rangle_\alpha = 0 = \widetilde{\alpha}_n^\mu |0\rangle_\alpha \quad \forall\, n > 0.
\end{split}
\label{eq:tensile_vac}
\end{equation}

As $\epsilon$ decreases, the vacuum evolves via a Bogoliubov transformation. In the \textit{tensionless limit} $\epsilon \to 0$, we find $\tanh\theta \to -1$.


Hence, the tensionless limit $\epsilon \to 0$ of the \textit{tensile closed string vacuum} $|0\rangle_\alpha$ flows to a state that is physically equivalent to the \textit{tensionless open string vacuum}, illustrating how open string degrees of freedom can emerge from closed strings in the null/tensionless regime.


Following \cite{Bagchi:2019cay}, Bogoliubov generator is expressed as 
\begin{equation}
G = i \sum_{n>0} \theta \left( C^\mu_{-n} R_\mu{}^\nu \widetilde{C}_{-n\,\nu} - \text{h.c.} \right).
\end{equation}
We have the transformed oscillators
\begin{equation}
e^{iG} C_n^\mu e^{-iG} = \cosh\theta\, C_n^\mu - \sinh\theta\, R^\mu{}_\nu \widetilde{C}_{-n}^\nu.
\end{equation}
Induced vacuum is 
\begin{equation}
\ket{0}_\alpha = e^{iG} \ket{0}_c = \mathcal{N} \prod_{n=1}^\infty \exp\left( \frac{\tanh\theta}{n} C^\mu_{-n} R_\mu{}^\nu \widetilde{C}_{-n\,\nu} \right) \ket{0}_c.
\end{equation}
We can recover the boundary state in the limit $\tanh\theta \to -1$
\begin{equation}
\ket{I_{\mathfrak{B}}} = \mathcal{N} \exp\left( -\sum_{n>0} \frac{1}{n} C^\mu_{-n} R_\mu{}^\nu \widetilde{C}_{-n\,\nu} \right) \ket{0}_c.
\end{equation}


\subsection{Physical interpretation of the gluing condition}

\subsection*{{$\mathfrak{B}=0$ case: Neumann gluing.}}
When the background Kalb--Ramond field is absent, the oscillator gluing condition simplifies to
\begin{equation}
\left( C_n^\mu + \widetilde{C}_{-n}^\mu \right) \ket{I_0} = 0.
\end{equation}
This corresponds to a Neumann boundary state, which describes an open string whose endpoints are free to propagate in all spacetime dimensions. For a detailed exposition of boundary states see \cite[cf. section \texttt{6.2.1 Boundary Conditions}]{Blumenhagen:2009zz}. Thus an open string is obtained by taking a tensionless limit of a closed string. An open string with Neumann boundary conditions along all spacetime directions is equivalent to a space-filling D25-brane in bosonic string theory. In this picture, what begins as a closed string theory can effectively give rise to-or be reinterpreted as-such a space-filling brane.
\subsection*{{Turning on a constant $\mathfrak{B}$.}}
When a constant background $\mathfrak{B}$-field is switched on, the gluing condition no longer equates $C_n$ and $\widetilde{C}_{-n}$ directly. Instead, it acquires the form
\begin{equation}
\left( C_n + R \widetilde{C}_{-n} \right) \ket{I_{\mathfrak{B}}} = 0,
\end{equation}
where $R$ is a constant matrix determined by the background. This eliminates one set of degrees of freedom, effectively inducing a boundary on the worldsheet and reorganizing the closed-string vacuum into the vacuum of an open string in the presence of the background $B$-field. This provides a physical mechanism by which a closed string dynamically transitions into an open string in the tensionless regime.

\section{Induced vacuum condition in the presence of $B$-field compactified on $T^{d}$}
\label{torustd}
We now consider a closed tensionless string compactified on $T^{d}$, propagating in background constant $B$-field \cite{Banerjee:2024fbi}. The anti-symmetric $B$-field has non-vanishing entries only along those compact directions.

Suppose the compactification radius of the $I$-th dimension is $R_I$. 
Compactifying $d$ directions means we identify the coordinates as
\begin{equation}
X^I \;\sim\; X^I + 2\pi \omega^I R_I, 
\qquad I = 1, \dots, d ,
\label{eq:compact_identification}
\end{equation}
so that the closed string coordinates satisfy
\begin{equation}
X^I(\sigma+2\pi,\tau) \;=\; X^I(\sigma,\tau) + 2\pi \omega^I R_I ,
\label{eq:compact_bc}
\end{equation}
with $\omega^I$ denoting integer winding numbers.

It is convenient to introduce dimensionless variables by rescaling
\begin{equation}
\widetilde{X}^I \equiv \frac{X^I}{R_I}.
\label{eq:rescaling}
\end{equation}
Under this redefinition the metric and $B$-field components in the compact directions transform as
\begin{align}
G_{IJ} &= 
\frac{\partial X^K}{\partial \widetilde{X}^I}\,
\frac{\partial X^L}{\partial \widetilde{X}^J}\, \eta_{KL}
= R_I R_J \eta_{IJ}, \nonumber \\
\mathbb{B}_{IJ} &= 
\frac{\partial X^K}{\partial \widetilde{X}^I}\,
\frac{\partial X^L}{\partial \widetilde{X}^J}\, \mathfrak{B}_{KL}
= R_I R_J \mathfrak{B}_{IJ},
\label{eq:rescaled_metric_B}
\end{align}
and the inverse metric reads
\begin{equation}
G^{IJ} = \frac{1}{R_I R_J}\, \eta^{IJ}.
\label{eq:inverse_metric}
\end{equation}
In the scaled coordinates the identification rule \eqref{eq:compact_identification} simplifies to
\begin{equation}
\widetilde{X}^I \;\sim\; \widetilde{X}^I + 2\pi \omega^I,
\label{eq:scaled_identification}
\end{equation}
which implies the periodicity condition
\begin{equation}
\widetilde{X}^I(\sigma+2\pi,\tau)-\widetilde{X}^I(\sigma,\tau) = 2\pi \omega^I .
\end{equation}
Recall, our gauge choice is, $V^{\alpha}=(v,0)$. Expressed in terms of $\widetilde{X}^I$, the gauge-fixed action becomes
\begin{equation}
S = \int d^2\xi \Big(
v^2\, \dot{\widetilde{X}}^I \dot{\widetilde{X}}^J G_{IJ}
+ \frac{1}{2 \pi}\epsilon^{a b} \mathbb{B}_{IJ} \,\partial_a \widetilde{X}^I \partial_b \widetilde{X}^J
\Big).
\label{eq:rescaled_action}
\end{equation}
Note that this action looks similar to the action \eqref{eq:tensionless_action}, but now with the rescaled $\eta$, $\mathfrak{B}$ and $X^{I}$s. The equation of motion for $\widetilde{X}^I$ reads,
\begin{equation}
    \ddot{\widetilde{X}}^I = 0.
\end{equation}
This equation can be solved with the modified periodic boundary condition using the following mode expansion,
\begin{equation}
    \widetilde{X}^{I}(\tau, \sigma) = \widetilde{x}^{I} + \sqrt{\frac{c'}{2}} A_0^{I} \sigma + \sqrt{\frac{c'}{2}} B_0^{I} \tau + i \sqrt{\frac{c'}{2}} \sum_{n \neq 0} \frac{1}{n} \left( A_n^{I} - i n \tau B_n^{I} \right) e^{-i n \sigma}.
\end{equation}
The momentum conjugate to $\widetilde{X}^I$ is
\begin{equation}
\widetilde{\Pi}^I = \frac{1}{c'} \sum_{J=1}^d G_{IJ}\, \dot{\widetilde{X}}^J 
+ \sum_{J=1}^d \mathbb{B}_{IJ}\, \widetilde{X}'^J .
\label{eq:canonical_momentum}
\end{equation}
Averaging over $\sigma$ then gives the centre-of-mass momentum, which combines both Kaluza-Klein and winding contributions:
\begin{equation}
\widetilde{\pi}^I = \frac{1}{2\pi}\int_0^{2\pi} d\sigma\, \widetilde{\Pi}^I
= \sum_{J=1}^d K^J G_{IJ} + \sum_{J=1}^d \mathbb{B}_{IJ}\, \omega^J .
\label{eq:cm_momentum}
\end{equation}

Because the canonical momentum generates translations, the factor  
$\exp\!\left(i\sum_I \widetilde{\pi}^I \widetilde{X}^I\right)$ in the quantum mechanical wavefunction must also respect the periodicity \eqref{eq:scaled_identification}. 
This requires  
\begin{equation}
\omega^I \widetilde{\pi}^I \in \mathbb{Z},
\end{equation}
so that $\widetilde{\pi}^I$ is quantized
\begin{equation}
\widetilde{\pi}^I = k^I, 
\qquad k^I \in \mathbb{Z}.
\label{eq:momentum_quantization}
\end{equation}
We again rewrite this expression in more convenient way \cite{Banerjee:2024fbi,Banerjee:2023ekd}, by defining
\begin{equation}
\widetilde{X}^I = \sqrt{\tfrac{c'}{2}}\, Y^I .
\label{eq:Y_field}
\end{equation}
The mode expansion of $Y^I$ is
\begin{equation}
Y^I = y^I + A^I_0 \sigma + B^I_0 \tau
+ i \sum_{n\neq 0} \frac{1}{n}\,(A^I_n - in\tau B^I_n)\, e^{-in\sigma}.
\label{eq:Y_expansion}
\end{equation}
Here, $\widetilde{x}^{I} =\sqrt{\frac{c'}{2}} y^{I}$. Compactification ensures that both zero modes are present. The $A^I_0$ and $B^I_0$ can be calculated as,
\begin{equation}
(B_0)^I = \sqrt{2c'} \Big(k^I - \sum_{J=1}^d B_{IJ}\, \omega^J\Big), 
\qquad
A^I_0 = \sqrt{\tfrac{2}{c'}}\, \omega^I .
\label{eq:zero_modes}
\end{equation}
Alternatively, one may decompose $Y^I$ into `left'- and `right'-movers in terms of the oscillators $C_n$ and ${\widetilde{C}_n}$ as
\begin{align}
Y^I_L &= y^I_L + k^I_L (\tau+\sigma)
+ i \sum_{n\neq0}\frac{1}{n}\,(C^I_n - in\tau C^I_n)\, e^{-in\sigma}, \nonumber \\
Y^I_R &= y^I_R + k^I_R (\tau-\sigma)
+ i \sum_{n\neq0}\frac{1}{n}\,(\widetilde{C}^I_n - in\tau \widetilde{C}^I_n)\, e^{in\sigma}.
\label{eq:Y_left_right}
\end{align}
The coefficients $k^I_{L,R}$ represent the `left'- and `right'-moving momenta of the string in compact directions. 
Imposing the identifications \eqref{eq:compact_identification} and \eqref{eq:momentum_quantization} yields
\begin{equation}
(k^I)_{L,R} = \frac{1}{\sqrt{2}}
\left[
\sqrt{c'} \Big(k^I - \sum_{J=1}^d \mathbb{B}_{IJ}\,\omega^J\Big)
\;\pm\; \frac{1}{\sqrt{c'}} \sum_{J=1}^d G_{IJ}\,\omega^J
\right].
\label{eq:LR_momenta}
\end{equation}

\subsection*{Gluing conditions}
As discussed in section \ref{sec:boundary_conditions}, the boundary conditions now reads as,
\begin{equation}
\partial_\tau \widetilde{X}_I + \tfrac{1}{2\pi}\,\mathbb{B}_{IJ}\,\partial_\sigma \widetilde{X}^J \bigg|_{\partial\Sigma} = 0.
\end{equation}
We choose $v=1$.

We now use the mode expansion $Y^I=Y_L^I+Y_R^I$ which are related to $\widetilde{X}^I$ via equation \eqref{eq:Y_field}. In terms of $Y^I$ the above equation can be rewritten as,
\begin{equation}
\partial_\tau Y_I + \tfrac{1}{2\pi}\,\mathbb{B}_{IJ}\,\partial_\sigma Y^J \bigg|_{\partial\Sigma} = 0.
\end{equation}
The $\tau$-derivative is
\begin{equation}
\partial_\tau Y^I = k_L^I + k_R^I + \sum_{n \neq 0} (C_n^I-i n \tau C_n^I) e^{-i n \sigma}+ \sum_{n \neq 0} (\widetilde{C}_n^I-i n \tau \widetilde{C}_n^I) e^{i n \sigma},
\end{equation}
and the $\sigma$-derivative is
\begin{equation}
\partial_\sigma Y^I = k_L^I - k_R^I + \sum_{n \neq 0} (C_n^I-i n \tau C_n^I) e^{-i n \sigma}+ \sum_{n \neq 0} (\widetilde{C}_n^I-i n \tau \widetilde{C}_n^I) e^{i n \sigma},
\end{equation}
Now, inserting these into the boundary condition, we get the following\\

(i) for the zero mode:
\begin{equation}
    k_I+\tfrac{1}{2 \pi} \mathbb{B}_{IJ} \kappa^J=0,
\end{equation}
where $k^I=(k_L^I+k_R^I)$ and $\kappa^I=(k_L^I-k_R^I)$.\\

(ii) for oscillator/non-zero ($n \neq 0$) modes: we extract the coefficient of $e^{-i n \sigma}$ and we get
\begin{equation}
\Bigg[\left( G_{IJ} + \frac{1 - i n \tau}{2\pi} \mathbb{B}_{IJ} \right) C_n^J + \left( G_{IJ} - \frac{1 + i n \tau}{2\pi} \mathbb{B}_{IJ} \right) \widetilde{C}_{-n}^J \Bigg] = 0,
\end{equation}
$G_{IJ}$ and $\mathbb{B}_{IJ}$ are defined in equation \eqref{eq:rescaled_metric_B}. Again, the above equation can be written in terms of $\eta_{IJ}$ and $\mathfrak{B}_{IJ}$ and this gives the same form as in \eqref{eq:main0} with $\mu=I$, i.e.,
\begin{equation}
\Bigg[\left( \eta_{IJ} + \frac{1 - i n \tau}{2\pi} \mathfrak{B}_{IJ} \right) C_n^J + \left( \eta_{IJ} - \frac{1 + i n \tau}{2\pi} \mathfrak{B}_{IJ} \right) \widetilde{C}_{-n}^J \Bigg] = 0.
\end{equation}
The induced vacuum $|I_\mathfrak{B}\rangle$ must satisfy this operator equation as a constraint. We have 
\begin{equation} \label{eq:main1}
\Bigg[\left( \eta_{IJ} + \frac{1 - i n \tau}{2\pi} \mathfrak{B}_{IJ} \right) C_n^J + \left( \eta_{IJ} - \frac{1 + i n \tau}{2\pi} \mathfrak{B}_{IJ} \right) \widetilde{C}_{-n}^J \Bigg] |I_\mathfrak{B}\rangle= 0.
\end{equation}

\subsection*{Gluing matrix}
We take the tensionless limit (worldsheet Carroll limit) by scaling $\tau \to \epsilon \tau$ and sending $\epsilon \to 0$. We get,
\begin{equation}
\frac{1 \mp i n \tau}{2\pi} \longrightarrow \frac{1}{2\pi}=a \quad(\text{say}).
\end{equation}
We finally have
 \begin{equation}
 \big(C_n + R\,\widetilde{C}_{-n}\big)\ket{I_{\mathfrak{B}}} = 0 \qquad (n > 0),
 \end{equation}
with the gluing matrix $R$ given by
\begin{equation}
R = (\eta + a \mathfrak{B})^{-1} (\eta - a \mathfrak{B}).
\end{equation}
The matrix $R$ acts on the indices of the oscillator modes. Its components are
\begin{equation}
R^I{}_J = \left[ (\eta + a \mathfrak{B})^{-1} (\eta - a \mathfrak{B}) \right]^I{}_J.
\end{equation}
The state $\ket{I_{\mathfrak{B}}}$ given by 
\begin{equation}
\begin{split}
\ket{I_{\mathfrak{B}}} &= \mathcal{N}\text{exp}\Bigg[{-\sum_{n>0}\frac{1}{n}\;C_{-n}^{I}\,R_{I}{}^{J}\,\widetilde{C}_{-n\,J}}\Bigg] \ket{0}_c.
\end{split}
\end{equation}
As $\mathfrak{B} \to 0$, $R \to 1$, so the state becomes
\begin{equation}
\ket{I_{\mathfrak{B}}} \xrightarrow{\mathfrak{B} \to 0} \mathcal{N} \exp\!\left( -\sum_{n>0} \frac{1}{n} C_{-n} \cdot \widetilde{C}_{-n} \right) \ket{0}_c,
\end{equation}
which is precisely the Neumann boundary state. 

\subsection*{Bogoliubov generator}
Bogoliubov generator is expressed as 
\begin{equation}
G = i \sum_{n>0} \theta \left( C^I_{-n} R_I{}^J \widetilde{C}_{-n\,J} - \text{h.c.} \right).
\end{equation}
We have the transformed oscillators
\begin{equation}
e^{iG} C_n^I e^{-iG} = \cosh\theta\, C_n^I - \sinh\theta\, R^I{}_J \widetilde{C}_{-n}^J.
\end{equation}
Induced vacuum is 
\begin{equation}
\ket{0}_\alpha = e^{iG} \ket{0}_c = \mathcal{N} \prod_{n=1}^\infty \exp\left( \frac{\tanh\theta}{n} C^I_{-n} R_I{}^J \widetilde{C}_{-n\,J} \right) \ket{0}_c.
\end{equation}
We can recover the boundary state in the limit $\tanh\theta \to -1$
\begin{equation}
\ket{I_\mathfrak{B}} = \mathcal{N} \exp\left( -\sum_{n>0} \frac{1}{n} C^I_{-n} R_I{}^J \widetilde{C}_{-n\,J} \right) \ket{0}_c.
\end{equation}

\subsection*{Bose--Einstein-like condensation on the worldsheet}
One of the most striking features of the tensionless strings is the collapse of excited tensile perturbative states onto the induced vacuum, a process that has been interpreted as a form of Bose--Einstein-like condensation \cite{Bagchi:2019cay}. In the absence of compactification or with trivial winding, oscillator excitations lose their individuality and collectively condense into the degenerate ground state. We now revisit this phenomenon in the presence of a constant Kalb--Ramond background.

Consider a physical tensile perturbative state,
\begin{equation}
    |\zeta_n\rangle = \rho_{\mu\nu} \, \alpha_{-n}^{\mu} \, \widetilde{\alpha}_{-n-\ell}^{\nu} \, |0,0,k^{\mu},k_{I},\omega^{I}\rangle_{\alpha} \,.
\end{equation}
where $\rho_{\mu\nu}$ is polarization tensor, $k_I$
are quantized momenta and $\omega^I$ are winding numbers.
The level matching condition for the tensile string with $d$ compactified directions is,
\begin{equation}
    \widetilde{N} - N = \sum_{I} k_{I} \omega^{I} \equiv \ell 
\end{equation}

Now, we want to understand the evolution of the state $|0,0,k^{\mu},k_{I},\omega^{I}\rangle_{\alpha}$ as we take tensionless limit ($\epsilon \to 0$). Around tensionless limit the tensile string vacuum can be perturbatively expanded in terms of $\epsilon$ in following way,
\begin{equation}
  |0,0,k^{\mu},k_{I},\omega^{I}\rangle_{\alpha}=|0\rangle_I + \epsilon \, |I_1\rangle + \epsilon^2 \, |I_2\rangle + \cdots  .
\end{equation}
Recalling the definition of tensile vacua
\begin{equation}
\alpha_n |0\rangle_\alpha = \widetilde{\alpha}_n |0\rangle_\alpha = 0 \quad \forall\, n > 0.
\end{equation}
If one writes down the tensile oscillators $\alpha_n$, $\widetilde{\alpha}_n$ in terms of tensionless oscillators $A_n$ and $B_n$, one can finally arrive at the following conditions (depends crucially on the value of $\ell$) in the $\epsilon \to 0$ limit:
\subsection*{{Case $\ell=0$: condensation.}}
The background $B_{IJ}$ only enters the level-matching condition through the combination 
$(k_I - B_{IJ}\,\omega^J)$. However, the antisymmetry of $B_{IJ}$ ensures that the bilinear 
form $B_{IJ}\,\omega^I \omega^J$ vanishes. Consequently, the condition $\ell=0$ remains unaltered 
by the $B$-field. In this case the perturbative state collapses onto the induced vacuum,
\begin{equation}
|\zeta_n\rangle \;\longrightarrow\; 2n\eta^{\mu \nu} \rho_{\mu \nu}\,|0,0,k^\mu,k_I,\omega^I\rangle_I 
\label{eq:BECvac}.
\end{equation}
This represents ``Bose--Einstein condensation'' 
of excitations into the degenerate ground state, exactly as in the $B=0$ theory. Here the condensation means \textit{condensation in mode space}.
\subsection*{{Case $\ell>0$: unphysical.}}
As before, the states with positive mismatch between left and right movers are projected out 
by the residual BMS constraints. The Kalb--Ramond background does not rescue these states, 
so they remain unphysical.
 
\subsection*{{Case $\ell<0$: partial survival.}}
If $n \ge |\ell|$, the state is again unphysical. If $n < |\ell|$, the state vanishes in the limit. 
Thus the condensation channel is completely obstructed once $\ell \neq 0$.

\medskip

In summary, the presence of a constant Kalb--Ramond field does not modify the essential 
mechanism of Bose--Einstein-like condensation in the induced vacuum sector. The antisymmetry of 
$B_{IJ}$ ensures that the condensation criterion remains governed by the simple momentum-winding 
pairing 
\begin{equation}
\ell = \sum_I k_I \,\omega^I .
\end{equation}
The $B$-field only reshuffles the definitions of the effective lattice vectors $(k_I,\omega^I)$, 
but once the level-matching condition is satisfied, the condensation into the induced vacuum 
proceeds identically.

\section{Symplectic derivation of noncommutative parameter: tensile strings}\label{sec:symplectic_tensile}
We now come to the other goal of this paper: to understand noncommutative picture for tensionless strings. In the tensionless regime, the worldsheet metric becomes degenerate, and the usual notions of left- and right-moving oscillators no longer exist. The worldsheet dynamics reduces to a first-order, ultralocal system rather than a two-dimensional conformal field theory. Because of this degeneration, the standard operator or path-integral approach
to compute the open-string two-point function cannot be applied; there is no nontrivial propagator or oscillator algebra from which one could extract a commutator. The standard operator approach \cite{Seiberg:1999vs} for computing noncommutative parameter using 2-pt function in tensile theory is discussed in appendix \ref{appendixB} and computation of 2-pt function for tensioless strings is discussed in appendix \ref{appendixC}. The symplectic formalism does not rely on mode expansions or worldsheet conformal structure, it 
operates at the classical level of action and identifies the boundary phase-space structure through the symplectic two-form. Because of this, we use the symplectic approach for the tensionless case, where the canonical kinetic term collapses and only the boundary $B$-term survives. In this limit, the noncommutative structure can still be derived consistently by inverting the boundary symplectic form, even though the usual 2d CFT machinery is no longer available. \\

 Before directly going into the tensionless case, in this section we plan to discuss the computation of the noncommutative parameter $\Theta^{ij}$ for tensile open strings using the symplectic 2-form of the theory. Consider an open tensile bosonic string propagating in a flat $(g, B)$ background, the Polyakov action in conformal gauge is
\begin{equation}
S = \frac{1}{4\pi\alpha'} \int d\tau d\sigma 
\left[
g_{ij}(\dot{X}^i\dot{X}^j - X^{\prime i}X^{\prime j})
+ 2(2\pi\alpha') B_{ij} \dot{X}^i X^{\prime j}
\right],
\label{eq:tensile_action}
\end{equation}
where dots and primes denote $\partial_\tau$ and $\partial_\sigma$ respectively.

Variation of the action yields the equation of motion for the free fields and the boundary condition at the boundary $\sigma=0,\pi$ is,
\begin{align}
g_{ij}X^{\prime j} + 2\pi\alpha' B_{ij}\dot{X}^j &= 0. \label{eq:tensile_bdy}
\end{align} 

The canonical momentum can be calculated as,
\begin{align}
\Pi_i &= \frac{1}{2\pi\alpha'} g_{ij}\dot{X}^j +  B_{ij} X^{\prime j}.
\label{eq:canonical_momenta}
\end{align} 

\subsection{The symplectic form}

In the canonical formulation, the coefficient of the time derivative in the
Lagrangian density defines the (pre-)symplectic one-form on the space of
fields \cite{Harlow:2019yfa}. The analogue of the mechanical relation $\theta = p_i\, \delta q^i$
is therefore
\begin{equation}
\theta_{\Sigma}
   = \int_{\Sigma} d\sigma\, \Pi_i\, \delta X^i,
\label{eq:symplectic-potential}
\end{equation}
which we refer to as the \emph{(pre-)symplectic potential} on an equal-time
slice $\Sigma$.  Its exterior derivative in field space,
$\Omega_{\Sigma} = \delta \theta_{\Sigma}
   = \int d\sigma\, \delta \Pi_i \wedge \delta X^i$,
defines the bulk symplectic form \cite{Chu:1998qz}.

For an open string, however, the boundary condition
\eqref{eq:tensile_bdy} constrains the variations of $\dot{X}$ and
$X'$ at the endpoints, so the independent physical degrees of freedom live
on the boundary phase space.
The relevant symplectic structure is obtained by restricting
\eqref{eq:symplectic-potential} to the boundary constraint surface,
where $\dot{X}$ is eliminated in favour of $X'$ (or vice versa) using
\eqref{eq:tensile_bdy}.
This restriction defines the \emph{boundary symplectic potential}
\begin{equation}
\theta_{\partial\Sigma}
   = \int_{\partial\Sigma} d\sigma\, \Pi_i\, \delta X^i,
\label{eq:boundary-potential}
\end{equation}
evaluated on the boundary condition
$g_{ij} X^{\prime j} + 2\pi\alpha' B_{ij} \dot{X}^j = 0$.
Taking its exterior derivative gives the corresponding boundary
symplectic form,
$\Omega_{\partial\Sigma} = \delta \theta_{\partial\Sigma}$,
\begin{equation}
    \Omega_{\partial\Sigma}
   = \int d\sigma\, \delta \Pi_i \wedge \delta X^i.
\end{equation}
When $B=0$, this defines the standard Poisson bracket for the fields $X^\mu$ and $\Pi^\mu$,
\begin{equation}
\begin{aligned}
\big\{\,X^{\mu}(\tau, \sigma),\, \Pi_{\nu}(\tau, \sigma')\,\big\}_{\text{P.B.}} 
   &= \delta^{\mu}_{\nu}\, \delta(\sigma - \sigma'), \\[4pt]
\big\{\,\Pi_{\mu}(\tau, \sigma),\, \Pi_{\nu}(\tau, \sigma')\,\big\}_{\text{P.B.}} 
   &= 0, \\[4pt]
\big\{\,X^{\mu}(\tau, \sigma),\, X^{\nu}(\tau, \sigma')\,\big\}_{\text{P.B.}} 
   &= 0.
\end{aligned}
\label{eq:canonical-poisson}
\end{equation}
These are the standard equal-time commutation relations for free fields. However, in the presence of $B$-fields, these are no longer valid; noncommutativity comes into play.

Equation \eqref{eq:tensile_bdy} can be written in matrix form as,
\begin{align}
\mathbf{g} X^{\prime } + 2\pi\alpha' \mathbf{B}\dot{X} &= 0. \label{eq:tensile_bdy_matrixform}
\end{align} 
We now write $X^{\prime } $ in terms of $\dot{X}$ and substitute it to the canonical momenta. This gives,
\begin{align}
    \Pi = \frac{1}{2\pi\alpha'}\, \mathbf{g} \dot{X} + \mathbf{B} X' 
= \left[ \frac{1}{2\pi\alpha'}\, \mathbf{g} - 2\pi\alpha' \,\mathbf{B} \mathbf{g}^{-1} \mathbf{B} \right] \dot{X}= \mathbf{A} \dot{X},
\end{align}
where $\mathbf{A}:=\frac{1}{2\pi\alpha'}\, \mathbf{g} - 2\pi\alpha' \,\mathbf{B} \mathbf{g}^{-1} \mathbf{B}$.\\

With this notation, the equal-time symplectic potential on the boundary can be written as,
\begin{equation}
\theta_{\partial \Sigma} = \int d\sigma \; \Pi_i \, \delta X^i 
= \int d\sigma \; \dot{X}^T \mathbf{A} \, \delta X,
\end{equation}
matrix multiplication is implied: 
$\dot{X}^T \mathbf{A} \, \delta X \equiv \dot{X}^i A_{ij} \delta X^j$. Varying this once more gives the (pre-)symplectic form,
\begin{equation}
\Omega_{\partial \Sigma} = \delta \theta_{\partial \Sigma} 
= \int d\sigma \; \bigl( \delta\dot{X}^T \mathbf{A} \, \delta X \bigr).
\end{equation}
Integrating by parts in $\tau$ (on the equal-time slice; boundary terms vanish) and antisymmetrizing (since $\delta X \wedge \delta X$ is antisymmetric), we obtain
\begin{equation}
\begin{split}
\Omega_{\partial \Sigma} 
&= \frac{1}{2} \int d\sigma \left( \delta\dot{X}^T \mathbf{A} \, \delta X - \delta X^T \mathbf{A} \, \delta\dot{X} \right) \\
&= \frac{1}{2} \int d\sigma \; \delta X^T \bigl( \overleftarrow{\partial_\tau} \mathbf{A} - \mathbf{A} \overrightarrow{\partial_\tau} \bigr) \delta X.
\end{split}
\end{equation}
Since $\mathbf{A}$ is constant (i.e. the background fields $g_{ij}$ and $B_{ij}$ are constant), $\partial_\tau \mathbf{A} = 0$, and this reduces to
\begin{equation}
\Omega_{\partial \Sigma} = \int d\sigma \; \delta X^T \mathbf{A} \, \delta\dot{X}
= \frac{1}{2} \int d\sigma \; \delta X^T \mathbf{A} \, \overleftrightarrow{\partial_\tau} \delta X,
\end{equation}
where the bidirectional derivative is defined by
\begin{equation}
\delta X^T \mathbf{A} \, \overleftrightarrow{\partial_\tau} \delta X 
\;\equiv\; \delta X^T \mathbf{A} \, \partial_\tau \delta X - (\partial_\tau \delta X)^T \mathbf{A} \, \delta X.
\end{equation}
A more convenient way is to rewrite    $\mathbf{A}$ in terms of the standard $\mathbf{E}$ and $\mathbf{E}^T$ matrices
\begin{equation}
\mathbf{E} := \mathbf{g} + 2\pi\alpha'\, \mathbf{B}, \qquad
\mathbf{E}^T :=\mathbf{g} - 2\pi\alpha'\, \mathbf{B},
\end{equation}
where $\mathbf{g}$ is a symmetric metric ($\mathbf{g}^T = \mathbf{g}$) and $\mathbf{B}$ is antisymmetric ($\mathbf{B}^T = -\mathbf{B}$), so $\mathbf{E}^T$ is indeed the transpose of $\mathbf{E}$. Then one has the algebraic identity
\begin{equation}
\mathbf{A} = \frac{1}{2\pi\alpha'}\, \mathbf{E} \, \mathbf{g}^{-1} \, \mathbf{E}^T.
\end{equation}
The (pre-)symplectic 2-form can now be written as,
\begin{equation}
\Omega_{\partial \Sigma} = \frac{1}{4\pi\alpha'} \int d\sigma \; \delta{X}^T \, (\mathbf{E} \, \mathbf{g}^{-1} \, \mathbf{E}^T) \, \overleftrightarrow{\partial_\tau} \delta X=\frac{1}{4\pi\alpha'} \int d\sigma \; \delta{X}^T \, \mathbf{M} \, \overleftrightarrow{\partial_\tau} \delta X,
\label{eq:defineM}
\end{equation}
where $\mathbf{M}=\mathbf{E} \, \mathbf{g}^{-1} \, \mathbf{E}^T$. We now use the boundary condition once to write $\dot{X}$ in terms of $X^{\prime}$,
\begin{equation}
\delta X' = -2\pi\alpha'\, \mathbf{g}^{-1} \mathbf{B}\, \delta\dot{X} \quad \Rightarrow \quad \delta\dot{X} = -\frac{1}{2\pi\alpha'}\, \mathbf{B}^{-1} \mathbf{g}\, \delta X'.
\end{equation}
Substituting this in equation \eqref{eq:defineM} gives the following,
\begin{equation}
\Omega_{\partial\Sigma} = \frac{1}{2\,(2\pi\alpha')^2} \int d\sigma \left[ \delta X'^T\, \mathbf{g}\, \mathbf{B}^{-1} \mathbf{M} \delta X \;+\; \delta X^T \mathbf{M} \mathbf{B}^{-1} \mathbf{g}\, \delta X' \right].
\end{equation}

\noindent Since all background fields are constant, and the boundary variations vanish, we use
\begin{equation}
\int d\sigma\, \delta X'^{T} \mathbf{Q} \, \delta X
= -\int d\sigma\, \delta X^{T} \mathbf{Q}^{T}\, \delta X',
\label{eq:defineQ}
\end{equation}
for any constant matrix \(\mathbf{Q}\).
Take \(\mathbf{Q} = \mathbf{g} \mathbf{B}^{-1} \mathbf{M}\), then
\begin{equation}
\mathbf{Q}^{T} = \mathbf{M}^{T} (\mathbf{B}^{-1})^{T} \mathbf{g}^{T} = \mathbf{M} (-\mathbf{B}^{-1}) \mathbf{g} = -\mathbf{M} \mathbf{B}^{-1} \mathbf{g}.
\end{equation}
Hence,
\begin{equation}
\int \delta X'^{T} \mathbf{g} \mathbf{B}^{-1} \mathbf{M}\, \delta X
= \int \delta X^{T} \mathbf{M} \mathbf{B}^{-1} \mathbf{g}\, \delta X'.
\end{equation}
Substituting back into \eqref{eq:defineQ}, both terms in the integrand are identical, and the prefactor $1/2$ cancels. We thus obtain
\begin{equation}
\Omega_{\partial\Sigma}
= \frac{1}{(2\pi\alpha')^2}\int d\sigma\,
\delta X^{T} \mathbf{M} \mathbf{B}^{-1} \mathbf{g}\, \delta X'.
\end{equation}
We antisymmetrize the expression in \(\sigma\) by noting that only the antisymmetric part of \(\mathbf{Q}\) contributes
\begin{align}
\int d\sigma\, \delta X^{T} \mathbf{Q} \, \delta X'
&= -\frac{1}{2}\int d\sigma\, \delta X^{T} \bigl(\mathbf{Q} - \mathbf{Q}^{T}\bigr) \overleftrightarrow{\partial_\sigma} \delta X, \\
\mathbf{Q}^{T} &= \mathbf{g}^{T} (\mathbf{B}^{-1})^{T} \mathbf{M}^{T}
= \mathbf{g}\, (-\mathbf{B}^{-1})\, \mathbf{M}
= -\mathbf{g}\, \mathbf{B}^{-1} \mathbf{M},
\end{align}
where we used \(\mathbf{g}^{T} = \mathbf{g}\), \(\mathbf{B}^{T} = -\mathbf{B}\), and \(\mathbf{M}^{T} = \mathbf{M}\) (since \(\mathbf{M} = \mathbf{E}\, \mathbf{g}^{-1} \mathbf{E}^{T}\) is symmetric).

Therefore,
\begin{equation}
\mathbf{Q} - \mathbf{Q}^{T} = \mathbf{M}\, \mathbf{B}^{-1} \mathbf{g} + \mathbf{g}\, \mathbf{B}^{-1} \mathbf{M}.
\end{equation}
Inserting this into the previous line gives
\begin{equation}
\Omega_{\partial\Sigma}
= -\frac{1}{2\,(2\pi\alpha')^{2}}
\int d\sigma\,
\delta X^{T}\,
\bigl(\mathbf{M}\,\mathbf{B}^{-1}\mathbf{g} + \mathbf{g}\,\mathbf{B}^{-1}\mathbf{M}\bigr)
\overleftrightarrow{\partial_\sigma}
\delta X.
\label{eq:gBM}
\end{equation}
The linearized open-string boundary condition is
\begin{equation}
\mathbf{g}\,\delta X' + 2\pi\alpha'\, \mathbf{B}\,\delta \dot X = 0
\quad \Longrightarrow \quad
\overleftrightarrow{\partial_\sigma}
= -\,2\pi\alpha'\, \mathbf{g}^{-1} \mathbf{B}\, \overleftrightarrow{\partial_\tau}.
\end{equation}
Substitute this into \eqref{eq:gBM}. The two minus signs cancel, and one factor of \(2\pi\alpha'\) cancels from the denominator, yielding
\begin{align}
\Omega_{\partial\Sigma}
&= \frac{1}{4\pi\alpha'} \int d\sigma\,
\delta X^{T}
\bigl(\mathbf{M}\,\mathbf{B}^{-1}\mathbf{g} + \mathbf{g}\,\mathbf{B}^{-1}\mathbf{M}\bigr) \mathbf{g}^{-1} \mathbf{B}\,
\overleftrightarrow{\partial_\tau}
\delta X \nonumber \\
&= \frac{1}{4\pi\alpha'} \int d\sigma\,
\delta X^{T}
\underbrace{\bigl(\mathbf{M} + \mathbf{g}\,\mathbf{B}^{-1}\mathbf{M}\,\mathbf{g}^{-1}\mathbf{B}\bigr)}_{\mathbf{S}}
\overleftrightarrow{\partial_\tau}
\delta X.
\end{align}
Hence,
\begin{equation}
\Omega_{\partial\Sigma}
= \frac{1}{4\pi\alpha'}\int d\sigma\,
\delta X^{T} \mathbf{S}\, \overleftrightarrow{\partial_\tau}\, \delta X,
\qquad 
\mathbf{S} = \mathbf{M} + \mathbf{g}\,\mathbf{B}^{-1}\mathbf{M}\,\mathbf{g}^{-1}\mathbf{B}.
\label{SgBM}
\end{equation}
By definition, the equal-time symplectic form can be written as
\begin{equation}
\Omega_{\partial\Sigma}[\delta_1,\delta_2]
= \frac{1}{2}\int d\sigma
\Big(
\delta_1 X^{T} \mathbf{K}\, \delta_2 X
- \delta_2 X^{T} \mathbf{K}\, \delta_1 X
\Big)
= \int d\sigma\, \frac{1}{2}\,\delta X^{i}\wedge \delta X^{j}\, K_{ij}.
\end{equation}

\paragraph{Equal-time interpretation.}
In constructing the boundary symplectic form we always work on a fixed time slice 
$\Sigma=\Sigma_\tau$ of the worldsheet.  
This ``equal-time'' restriction is essential: the symplectic structure is a geometric
two-form on phase space and must therefore be defined using the values of the fields
at a single time.

The presymplectic potential and its exterior derivative are evaluated at fixed $\tau$,
\begin{equation}
\theta_{\Sigma_\tau}
=\int d\sigma\,\Pi_i(\tau,\sigma)\,\delta X^i(\tau,\sigma),
\qquad
\Omega_{\Sigma_\tau}=\delta\theta_{\Sigma_\tau}.
\end{equation}
Any expression involving the bidirectional time derivative,
\begin{equation}
\delta X^{T} S\,\overleftrightarrow{\partial_\tau}\,\delta X
=
\delta X^{T} S\,\partial_\tau\delta X
-
(\partial_\tau\delta X)^{T} S\,\delta X,
\end{equation}
is therefore interpreted purely as an antisymmetrization in field space.  
Because all variations are evaluated at the same value of~$\tau$, the actual time derivative
$\partial_\tau$ never acts on physical evolution; it only serves as a bookkeeping device
to extract the antisymmetric part of the kernel $S$.  
Thus we use the standard identity
\begin{equation}
\frac{1}{2}\int d\sigma\,
\delta X^{T} S\,\overleftrightarrow{\partial_\tau}\,\delta X
\;=\;
\int d\sigma\,\delta X^{T} S\,\delta X,
\end{equation}
which converts the ``equal-time'' antisymmetrized derivative into the wedge product
$\delta X \wedge \delta X$ defining the symplectic two-form.  
In this way the final symplectic structure is entirely algebraic, as required for a
well-defined Poisson bracket on the equal-time phase space.

\medskip

For constant backgrounds, the kernel $\mathbf{K}$ obtained in this way contains no
worldsheet derivatives and is entirely determined by the matrices $\mathbf{g}$, $\mathbf{B}$, and
$\mathbf{E}=\mathbf{g}+2\pi\alpha' \mathbf{B}$.
Using matrix manipulations with \(\mathbf{S}\) from \eqref{SgBM} and the definition \(\mathbf{M} = \mathbf{E}\, \mathbf{g}^{-1} \mathbf{E}^{T}\), one finds
\begin{equation}
\begin{split}
\mathbf{K} &= \frac{1}{4\pi\alpha'}\,(\mathbf{E}\, \mathbf{g}^{-1} \mathbf{E}^{T})^{-1}\,(2\pi\alpha'\, \mathbf{B})\\
&=\frac{1}{4\pi\alpha'} (\mathbf{S}\, \mathbf{B}^{-1} \mathbf{g} + \mathbf{g}\, \mathbf{B}^{-1} \mathbf{S} ) \\
&= \frac{1}{4\pi\alpha'}\, \mathbf{M}^{-1}\,(2\pi\alpha'\, \mathbf{B}),
\label{eq:K_A}
\end{split}
\end{equation}
which is antisymmetric because \(\mathbf{M}^{-1}\) is symmetric and \(\mathbf{B}\) is antisymmetric.


The symplectic 2-form takes the compact, algebraic form
\begin{equation}
\Omega_{\partial\Sigma}
= \frac{1}{4\pi\alpha'}\int d\sigma\,
\delta X^{T}
\Big[(\mathbf{E}\, \mathbf{g}^{-1}\mathbf{E}^{T})^{-1}(2\pi\alpha'\, \mathbf{B})\Big]
\delta X
= \int d\sigma\, \frac{1}{2}\,\delta X^{i}\wedge\delta X^{j}\, K_{ij}.
\end{equation}

\subsection{The noncommutative parameter}
The appearance of noncommutativity can be understood naturally from the symplectic structure of the theory. Given a symplectic two-form
\begin{equation}
\Omega
=
\frac12\,\Omega_{ij}\, d\xi^i \wedge d\xi^j,
\end{equation}
the corresponding Poisson structure is determined by the inverse symplectic matrix
\begin{equation}
\Omega_{ik}\Omega^{kj}
=
\delta_i^{\ j}.
\end{equation}
The Poisson bracket between two phase-space functions $f(\xi)$ and $g(\xi)$ is then defined as
\begin{equation}
\{f,g\}
=
\Omega^{ij}\,
\partial_i f\,\partial_j g.
\end{equation}
In particular, the coordinates satisfy
\begin{equation}
\{\xi^i,\xi^j\}
=
\Omega^{ij}.
\end{equation}
Upon quantization, the Poisson brackets are replaced by commutators,
\begin{equation}
[\xi^i,\xi^j]
=
i\,\Omega^{ij},
\end{equation}
so that the inverse symplectic matrix plays the role of the noncommutative parameter. We present a brief review about basics of symplectic form, Poission algebra, Moyal-$\star$ product in Appendix \ref{appendixA}.\\

In the present case, the boundary dynamics induces the symplectic structure
\begin{equation}
\Omega
=
\frac12\,K_{ij}\, dX^i\wedge dX^j.
\end{equation}
From Eq.~\eqref{eq:K_A}, we have
\begin{equation}
\mathbf{K}
=
\frac{1}{4\pi\alpha'}
\left(
\mathbf{E}\, \mathbf{g}^{-1} \mathbf{E}^T
\right)^{-1}
(2\pi\alpha' \mathbf{B}).
\end{equation}
The Poisson tensor is the inverse of the symplectic matrix $\mathbf{K}$ \cite{Lee:1990nz,Nair:2024wyq},
\begin{equation}
\mathbf{\Theta}
=
\mathbf{K}^{-1}.
\end{equation}
Using the identity
\begin{equation}
\left(
\mathbf{E} \mathbf{g}^{-1} \mathbf{E}^T
\right)^{-1}
(2\pi\alpha' \mathbf{B})
=
-\,2\,\mathbf{E}^{-1}\mathbf{B}(\mathbf{E}^T)^{-1},
\end{equation}
we obtain
\begin{equation}
\mathbf{K}
=
-\frac12\,
\mathbf{E}^{-1}\mathbf{B}(\mathbf{E}^T)^{-1}.
\end{equation}
Therefore, the noncommutative parameter becomes
\begin{equation}
\mathbf{\Theta}
=
-(2\pi\alpha')^2
\left[
(\mathbf{g}+2\pi\alpha' \mathbf{B})^{-1}
\,\mathbf{B}\,
(\mathbf{g}-2\pi\alpha' \mathbf{B})^{-1}
\right].
\end{equation}
This is precisely the Seiberg--Witten noncommutative parameter \cite{Seiberg:1999vs}. Decomposing
\begin{equation}
(\mathbf{g}+2\pi\alpha' \mathbf{B})^{-1}
\end{equation}
into symmetric and antisymmetric parts reproduces the standard open-string metric
\begin{equation}
\mathbf{G}
=
\mathbf{g}
-
(2\pi\alpha')^2
\mathbf{B} \mathbf{g}^{-1} \mathbf{B},
\end{equation}
together with the familiar boundary correlator structure of noncommutative open strings \cite{Seiberg:1999vs}.

In the large-$B$ limit \cite{Seiberg:1999vs}, the noncommutative parameter reduces to
\begin{equation}
\mathbf{\Theta}
\;\longrightarrow\;
\mathbf{B}^{-1},
\end{equation}
up to convention-dependent normalization factors.

\section{Noncommutativity in tensionless strings}
\label{sec:symplectic_tensionless}

We now derive the noncommutative parameter for the \emph{tensionless} (null) string in the
presence of a constant background $B$-field. In the tensile open string, noncommutativity is
usually extracted from the boundary two-point function of the worldsheet conformal field
theory. That route is no longer natural in the intrinsically tensionless theory, because the
Carrollian contraction degenerates the usual two-dimensional propagating structure of the
worldsheet. The more robust object is instead the boundary symplectic form.\\



We begin with the gauge-fixed ILST action in the presence of a constant background
$B$-field,
\begin{equation}
S=\int d\tau\, d\sigma\,
\left[
v^2\, g_{ij}\,\dot X^i \dot X^j
+\frac{1}{\pi}\, B_{ij}\,\dot X^i X'^j
\right],
\qquad
v^2=\frac{1}{2\pi c'}.
\label{eq:tensionless-action-rewrite}
\end{equation}
Here dots and primes denote derivatives with respect to $\tau$ and $\sigma$, respectively,
and both $g_{ij}$ and $B_{ij}$ are taken to be constant, with $B_{ij}=-B_{ji}$.


Varying $X^i \to X^i+\delta X^i$, one finds
\begin{equation}
\begin{split}
\delta S
&=
\int d\tau\, d\sigma\,
\left[
2v^2 g_{ij}\,\dot X^j\,\delta \dot X^i
+\frac{1}{\pi} B_{ij}\,\delta \dot X^i\, X'^j
+\frac{1}{\pi} B_{ij}\,\dot X^i\, \delta X'^j
\right].
\end{split}
\end{equation}
After integrating by parts in $\tau$ and $\sigma$, and using that $g_{ij}$ and $B_{ij}$ are
constant, the bulk variation becomes
\begin{equation}
\delta S_{\text{bulk}}
=
-\int d\tau\, d\sigma\,
\left[
2v^2 g_{ij}\,\ddot X^j
\right]\delta X^i.
\end{equation}
Hence the bulk equation of motion is simply
\begin{equation}
\ddot X^i=0.
\label{eq:eom-null}
\end{equation}
The coefficient of $\delta X^i$ on a fixed-$\tau$ slice is
\begin{equation}
\int d\sigma\,
\left[
2v^2 g_{ij}\,\dot X^j
+\frac{1}{\pi} B_{ij}\,X'^j
\right]\delta X^i.
\label{eq:tau-boundary-term}
\end{equation}
This is the quantity that determines the canonical 1-form and hence the symplectic
structure. It is therefore natural to identify the canonical momentum density as
\begin{equation}
\Pi_i
=
\frac{\partial \mathcal L}{\partial \dot X^i}
=
2v^2 g_{ij}\,\dot X^j
+\frac{1}{\pi} B_{ij}\,X'^j.
\label{eq:momentum-null}
\end{equation}
Using $v^2=1/(2\pi c')$, this can be written in matrix notation as
\begin{equation}
\Pi
=
\frac{1}{\pi}
\left(
\frac{1}{c'}\,\mathbf g\,\dot X
+\mathbf B\,X'
\right).
\label{eq:momentum-null-matrix}
\end{equation}
Now, the tensionless open-string sector is characterized by the mixed relation
\begin{equation}
\frac{1}{c'}\,\mathbf g\,\dot X+\mathbf B\,X'=0.
\label{eq:null-mixed-bc}
\end{equation}
This is the null-string analogue of the familiar mixed Neumann boundary condition in the
tensile theory \cite{Seiberg:1999vs},
\begin{equation}
g_{ij}\,\partial_n X^j + 2\pi \alpha' B_{ij}\,\partial_t X^j =0.
\end{equation}
Equation \eqref{eq:null-mixed-bc} has an important consequence
\begin{equation}
\Pi_i\big|_{\partial\Sigma}=0.
\label{eq:Pi-zero}
\end{equation}

\subsection{Boundary symplectic form and intrinsic noncommutativity}

Let us now isolate the term responsible for the reduced boundary symplectic structure
\begin{equation}
\mathcal L_{\text{first-order}}
=
\frac{1}{\pi} B_{ij}\,\dot X^i X'^j.
\label{eq:first-order-lagrangian}
\end{equation}
From \eqref{eq:first-order-lagrangian}, one reads off the boundary symplectic potential in the
form
\begin{equation}
\theta_{\partial\Sigma}
=
\int d\sigma\; a_i(X)\,\delta X^i,
\qquad
a_i(X)=\frac{1}{\pi} B_{ij}\,X'^j.
\label{eq:theta-general}
\end{equation}
For constant $B$, and up to the usual freedom of adding field-space exact terms or total
$\sigma$-derivatives, one may choose the equivalent representative
\begin{equation}
\theta_{\partial\Sigma}
=
\frac{1}{2\pi}
B_{ij}\,X^i\,\delta X^j \Big|_{\partial\Sigma}.
\label{eq:theta-bdy-rewrite}
\end{equation}
Taking the exterior derivative in field space, we obtain the reduced boundary symplectic form
\begin{equation}
\Omega_{\partial\Sigma}
=
\delta \theta_{\partial\Sigma}
=
\frac{1}{2\pi}
B_{ij}\,\delta X^i \wedge \delta X^j\Bigg|_{\partial\Sigma}.
\label{eq:omega-bdy-rewrite}
\end{equation}
Several important features are immediate.

First, \eqref{eq:omega-bdy-rewrite} is purely algebraic in the boundary fields: it contains no
worldsheet derivatives and depends only on $B_{ij}$. This makes precise the statement that,
in the reduced boundary theory, the antisymmetric background alone carries the symplectic
data.

Second, the induced symplectic form is independent of both $c'$ and $g_{ij}$. This is the
sharp distinction from the tensile theory. In the latter, the noncommutative parameter is a
combined effect of the metric and the $B$-field. In the strict tensionless theory, by contrast,
the metric contribution drops out of the reduced boundary symplectic structure, and the
antisymmetric background is the unique surviving source of noncommutativity.

Third, if $B_{ij}$ is invertible on the relevant target-space subspace, then the symplectic
kernel is
\begin{equation}
\omega_{ij}(\sigma,\sigma')
=
\frac{1}{2\pi} B_{ij}\,\delta(\sigma-\sigma').
\end{equation}
The equal-time Poisson brackets on the reduced boundary phase space are therefore
\begin{equation}
\{X^i(\sigma),X^j(\sigma')\}_{\partial\Sigma}
=
2\pi\,B^{ij}\,\delta(\sigma-\sigma').
\label{eq:PB-null}
\end{equation}
Upon quantization, this becomes
\begin{equation}
[X^i(\sigma),X^j(\sigma')]
=
i\,2\pi\,B^{ij}\,\delta(\sigma-\sigma').
\label{eq:comm-null}
\end{equation}
Thus the noncommutative parameter of the tensionless string is
\begin{equation}
\Theta^{ij}_{\text{tensionless}}
=
2\pi\,B^{ij},
\label{eq:Theta-null}
\end{equation}
on the subspace where $\mathbf{B}$ is non-degenerate. If $\mathbf{B}$ has a nontrivial kernel, the inversion is
understood on its support, and the null directions remain commuting.

Equation \eqref{eq:Theta-null} is the intrinsic null-string counterpart of the familiar
Seiberg--Witten noncommutativity parameter. The resemblance in form is important, but so
is the difference in origin. In the tensile theory, noncommutativity emerges from the boundary
operator algebra of a conventional two-dimensional conformal field theory. Here, instead, it
arises because the Carrollian limit collapses the standard coordinate--momentum structure on
the boundary, leaving the first-order $B$-coupling as the unique symplectic datum. In this
precise sense, the noncommutative geometry of the tensionless string is not merely inherited
from the tensile theory; it is encoded directly in the reduced boundary phase space of the
intrinsically Carrollian worldsheet.

\subsection{$\Theta_{\text{tensile}} \to \Theta_{\text{tensionless}}$}

The two noncommutative parameters for tensile and tensionless theory presented above are structurally similar, yet they describe
two distinct physical regimes of the string worldsheet.

In the \emph{tensile} case, the worldsheet theory is a genuine two-dimensional
conformal field theory with a non-degenerate induced metric. 
Both the metric term and the $B$-term contribute to the canonical structure.
The full boundary symplectic kernel in that case is
\begin{equation}
\mathbf{K}^{(\text{tensile})}
= \frac{1}{4\pi\alpha'}\,\mathbf{M}^{-1}\,(2\pi\alpha'\,\mathbf{B}),
\qquad 
\mathbf{M} = \mathbf{E}\,\mathbf{g}^{-1}\,\mathbf{E}^{T},\quad 
\mathbf{E} = \mathbf{g} + 2\pi\alpha'\,\mathbf{B}.
\end{equation}
Inverting the symplectic kernel gives the Poisson tensor
\begin{equation}
\boldsymbol{\Theta}_{\text{tensile}}
= -\, (2\pi\alpha')^{2}
\Big[
(\mathbf{g} + 2\pi\alpha'\mathbf{B})^{-1}\,
\mathbf{B}\,
(\mathbf{g} - 2\pi\alpha'\mathbf{B})^{-1}
\Big].
\label{eq:tensileTheta}
\end{equation}
This expression is the standard Seiberg--Witten formula for the
noncommutative parameter \cite{Seiberg:1999vs}.
It contains both symmetric and antisymmetric contributions, the former giving
the open-string metric
\(
\mathbf{G} = \mathbf{g} - (2\pi\alpha')^{2}\,\mathbf{B}\mathbf{g}^{-1}\mathbf{B}.
\)

\paragraph{Tensionless reduction.}
The \emph{tensionless} (or Carrollian) string corresponds to the regime
\begin{equation}
T \sim \frac{1}{2\pi\alpha'} \longrightarrow 0,
\qquad \text{or equivalently } \alpha' \longrightarrow \infty.
\end{equation}
In this limit, the coefficient of the metric term in the action vanishes,
and the dynamics becomes first order in time derivatives.

To extract the corresponding limit of \eqref{eq:tensileTheta}, note that
the combination $2\pi\alpha' \mathbf{B}$ appears ubiquitously.
Define
\begin{equation}
\mathbf{F} := 2\pi\alpha' \mathbf{B},
\end{equation}
and take $\alpha' \to \infty$ keeping $\mathbf{F}$ finite and large.
Then
\[
\mathbf{E} = \mathbf{g} + \mathbf{F}
\simeq \mathbf{F}, \qquad
\mathbf{E}^{T} = \mathbf{g} - \mathbf{F} \simeq -\,\mathbf{F}.
\]
Substituting into \eqref{eq:tensileTheta},
\begin{align}
\boldsymbol{\Theta}_{\text{tensile}}
&\simeq -\, (2\pi\alpha')^{2}\,
\Big[
\mathbf{F}^{-1}\,\mathbf{B}\,(-\mathbf{F}^{-1})
\Big]
= (2\pi\alpha')^{2}\,\mathbf{F}^{-1}\mathbf{B}\mathbf{F}^{-1}.
\end{align}
Using $\mathbf{F} = 2\pi\alpha'\mathbf{B}$, we find
\begin{equation}
\boldsymbol{\Theta}_{\text{tensile}}
\;\xrightarrow[\alpha' \to \infty]{}\;
\mathbf{B}^{-1}.
\label{eq:tensile-to-tensionless}
\end{equation}
This is precisely the noncommutativity tensor
derived directly from the tensionless analysis,
\(
\Theta^{ij}_{\text{tensionless}} = B^{ij}.
\)

\paragraph{Physical interpretation.}
The equivalence \eqref{eq:tensile-to-tensionless} can be interpreted in two different ways:

\begin{enumerate}
    \item \emph{In the tensile theory}, the limit $\alpha' \to \infty$
    (or equivalently vanishing tension $T \to 0$)
    suppresses the second-order metric term in the Polyakov action.
    The worldsheet metric becomes degenerate, and the first-order
    $B$-term dominates, turning the theory effectively into a
    topological boundary system.

    \item \emph{In the tensionless formulation}, one starts from the
    degenerate worldsheet from the outset. The phase-space structure
    is governed entirely by the $B$-term, and its inverse defines
    the Poisson brackets directly. No limiting procedure is required:
    noncommutativity is an intrinsic property of the null worldsheet.
\end{enumerate}
Therefore, the tensionless symplectic structure
\begin{equation}
\Omega_{\partial\Sigma}
= \frac{1}{2\pi} \int d\sigma\, B_{ij}\, \delta X^i \wedge \delta X^j,
\qquad
\Theta^{ij}_{\text{tensionless}} = B^{ij},
\end{equation}
is the natural Carrollian analogue of the tensile symplectic form
\begin{equation}
\Omega_{\partial\Sigma}
= \frac{1}{4\pi\alpha'} \int d\sigma\,
\delta X^T (\mathbf{E}\mathbf{g}^{-1}\mathbf{E}^T)^{-1}(2\pi\alpha'\mathbf{B})\,\delta X,
\end{equation}
\begin{equation}
    \Theta^{ij}_{\text{tensile}} 
= - (2\pi\alpha')^2 
\bigl[(\mathbf{g}+2\pi\alpha'\mathbf{B})^{-1}\mathbf{B}(\mathbf{g}-2\pi\alpha'\mathbf{B})^{-1}\bigr]^{ij}.
\end{equation}
The two forms match exactly in the large-$B$, zero-tension limit of tensile theory.
In other words,
\begin{equation}
\lim_{\substack{\alpha'\to \infty \\[2pt] 2\pi\alpha'\mathbf{B}\ \text{fixed}}}
\Theta^{ij}_{\text{tensile}}
= \Theta^{ij}_{\text{tensionless}} = B^{ij}.
\end{equation}

The tensionless string thus provides an \emph{intrinsic geometric realization}
of noncommutativity that in the tensile case emerges only after taking
the $\alpha' \to \infty$ limit.  
Both descriptions yield the same antisymmetric Poisson structure on the boundary phase space, but their origin differs:
in the tensile case, it is a limiting deformation of the full 2D CFT;
in the tensionless case, it is a built-in feature of the null (Carrollian)
worldsheet geometry.

\section{Conclusions and future directions}

In this work, we have presented a comprehensive analysis of the emergence of open strings from closed tensionless strings in the presence of a uniform Kalb--Ramond background and investigated how noncommutativity arises in this setting through a purely symplectic approach. 

In the first part, we established that the inclusion of a constant antisymmetric Kalb--Ramond background modifies the induced vacuum of the tensionless string, leading to a generalized gluing condition between modes. This deformation is encoded in a constant gluing matrix which smoothly interpolates between the standard Neumann condition and its background-deformed version. The resulting induced vacuum takes the form of a squeezed boundary state, manifesting the continuous evolution from the closed-string vacuum to an open-string-like configuration as the tension goes to zero. Furthermore, we extended this construction to the case of compactification on $T^d$, where the interplay between winding and momentum modes leads to a Kalb--Ramond background-field-dependent deformation of the induced vacuum, while the mechanism of Bose--Einstein condensation on the worldsheet remains robust and unaltered by the antisymmetric background.

\medskip

In the second part, we investigated the symplectic and geometric origin of the noncommutative structure associated with strings in constant $(g,B)$ backgrounds. Starting from first principles, we showed that both the tensile and the tensionless strings can be described within a unified symplectic framework. The boundary contribution to the symplectic potential encodes all the information about the noncommutative phase-space geometry. In the tensile regime, this reproduces the well-known Seiberg--Witten relation between the open-string metric $G_{ij}$, the antisymmetric parameter $\Theta^{ij}$, and the background $B_{ij}$. In the tensionless regime, the symplectic structure degenerates in the bulk but survives on the boundary, where it defines an intrinsic noncommutative geometry governed solely by the background $B$-field. Remarkably, the noncommutativity of the tensionless string arises not as a limit of the tensile theory, but as an intrinsic feature of the null (Carrollian) worldsheet.

In a recent study \cite{Das:2026efq}, we took an alternative approach to calculating the noncommutativity parameter by applying the covariant phase space formalism developed by Crnković, Witten, and Zuckerman~\cite{Crnkovic:1986ex,Zuckerman:1986vzu} (see also \cite{Crnkovic:1986,Crnkovic:1987tz,Grant:2005qc,Maoz:2005nk,Witten:1986qs}). Because this method builds the symplectic form directly from the action without relying on equal-time slicing, it seamlessly handled boundary effects introduced by a background $B$-field. We applied this framework to investigate open string noncommutativity. For standard tensile strings in a uniform Kalb-Ramond field, we demonstrated that the (pre)-symplectic current divided into a bulk kinetic component and an exact boundary piece, successfully yielding the Seiberg-Witten parameter. When we generalized this to intrinsically tensionless strings, we found that without background fields, the reduced phase space became degenerate and lacked a fundamental Poisson structure. However, introducing a constant Kalb-Ramond field caused the symplectic current to localize completely at the boundary. Consequently, the physical phase space existed solely on the boundary, granting the endpoint coordinates a noncommutative Poisson algebra. Similarly, adding a boundary gauge-field coupling produced a boundary symplectic form dictated by the effective Born-Infeld dynamics on the D-brane. Together, these findings offered a cohesive framework for understanding noncommutativity across both tensile and tensionless string regimes.
\subsection*{Future directions}

Our analysis opens several avenues for further research.

\paragraph{Flux backgrounds and the role of $\mathbf{\mathcal{F}}=B+F$ in the tensionless phase.}
A natural extension of our analysis is to incorporate a worldvolume gauge field $A_\mu$, which in the tensile string enters only through the gauge-invariant combination 
$\mathcal{F}=B+F$. Our results indicate that the emergent open string phase and its boundary symplectic structure behave as a Carrollian analogue of a brane. In this framework, turning on 
$F_{\mu\nu}$ amounts to the replacement $B \to \mathcal{F}$ in the boundary conditions, gluing matrix, and symplectic two-form. This would lead to a generalized noncommutative parameter 
$\Theta \sim \mathcal{F}^{-1}$, providing a tensionless counterpart of the familiar $B+F$ deformation in the Seiberg--Witten analysis. Understanding how gauge flux and bound-state 
structures arise directly from the null worldsheet, without relying on a pre-existing D-brane, would offer a novel perspective on flux backgrounds in the Carrollian regime and may reveal tensionless analogues of dielectric or Myers-type effects.

\paragraph{Supersymmetric generalization.}  
Incorporating worldsheet fermions $\psi^i$ and their conjugate momenta extends the symplectic form to a super-symplectic structure. Such an extension would allow one to study the deformation of the boundary superalgebra induced by the background field and to determine how the supersymmetry transformations act on the noncommutative phase space of the endpoints.


\paragraph{Slowly varying backgrounds.}
The analysis in this paper assumed constant $(g,B)$ fields, for which the boundary
symplectic kernel becomes purely algebraic.  
A natural extension is to treat slowly varying backgrounds with 
$\partial_k B_{ij}\neq 0$.  
In \cite{Seiberg:1999vs}, Seiberg and Witten showed that such variations
produce derivative corrections to the open-string metric, the noncommutativity
parameter.  
From the symplectic point of view, these corrections would manifest as new
$\partial B$ contributions to the boundary two-form, making the kernel no longer
purely algebraic.  
Understanding these corrections directly at the level of the boundary symplectic
structure could give a geometric derivation of the derivative-corrected
open-string data and may shed light on how noncommutativity behaves for
tensionless strings in inhomogeneous backgrounds.

\paragraph{Quantization and Moyal deformation.}
Our symplectic derivation naturally suggests a quantization scheme for the boundary phase space of tensionless strings. A detailed study of the associated Moyal $\star$-algebra \cite{Herbst:2001ai} and its physical interpretation could yield new insights into the quantum geometry of null strings.

\paragraph{World-volume actions for D-branes in the tensionless regime.}
Given that the induced vacuum behaves as an open string attached to a space-filling brane, it would be valuable to derive the world-volume actions for D-branes. A Carrollian analogue of the Born--Infeld action, particularly in the presence of a constant $B$-field, may clarify the emergence of gauge dynamics in the tensionless phase.


\paragraph{Higher-form backgrounds and generalized charges.}
Including RR fields or higher $p$-form backgrounds can test the robustness of the closed-to-open transition. Such setups may lead to tensionless analogues of dielectric branes or Myers-like effects \cite{Myers:1999ps} driven by worldsheet constraints.

\section*{Acknowledgements}
We would like to thank Mohammad M. Sheikh-Jabbari for useful comments on our paper. The work was presented by SD in the Quantum Fields and Strings Group Seminar at BIMSA. We thank Hossein Yavartanoo for the kind invitation. SD is supported by the Shuimu Tsinghua Scholar Program of Tsinghua University and the Beijing Natural Science Foundation of China Grant No.~IS25035. SM thanks the String Theory group of HRI for useful discussions.

\appendix

\section{Boundary state formalism}\label{appA}
Boundary states provide a closed-string description of D-branes and open-string boundary conditions. The basic idea is that a worldsheet with boundaries admits two equivalent interpretations related by worldsheet duality. In the open-string channel, the worldsheet coordinates are $(\tau,\sigma)$, where $\tau$ denotes propagation time and the boundaries are located at fixed $\sigma$, typically $\sigma=0,\pi$. Interchanging the roles of $\tau$ and $\sigma$,
\begin{equation}
(\sigma,\tau)_{\text{open}}
\longleftrightarrow
(\tau,\sigma)_{\text{closed}},
\end{equation}
the same cylinder can instead be viewed as a closed string propagating between two boundary states \cite{Blumenhagen:2009zz,Blumenhagen:2013fgp}. See figure \ref{fig:worldsheet_duality}.

\begin{figure}[H]
\centering

\begin{tikzpicture}[x=1cm,y=1cm,line join=round,line cap=round]


\fill[gray!25]
(-6,-2.2) -- (-4.3,-1.1) -- (-4.3,2.2) -- (-6,1.1) -- cycle;

\draw
(-6,-2.2) -- (-4.3,-1.1)
(-4.3,-1.1) -- (-4.3,2.2)
(-4.3,2.2) -- (-6,1.1)
(-6,1.1) -- (-6,-2.2);

\fill[gray!20]
(-1.8,-2.2) -- (-0.1,-1.1) -- (-0.1,2.2) -- (-1.8,1.1) -- cycle;

\draw
(-1.8,-2.2) -- (-0.1,-1.1)
(-0.1,-1.1) -- (-0.1,2.2)
(-0.1,2.2) -- (-1.8,1.1)
(-1.8,1.1) -- (-1.8,-2.2);

\shade[left color=blue!20,right color=blue!55,opacity=0.45]
(-4.9,-0.7) rectangle (-1.2,0.7);

\shade[inner color=blue!10,outer color=blue!50,opacity=0.6]
(-4.9,0) ellipse (0.42 and 0.72);

\draw (-4.9,0) ellipse (0.42 and 0.72);

\fill[gray!10] (-1.2,0) ellipse (0.42 and 0.72);
\draw (-1.2,0) ellipse (0.42 and 0.72);

\draw[->,thick]
(-5.30,-0.05)
.. controls (-3.8,0.35) and (-2.4,0.15)
.. (-1.55,0.02);

\node at (-3.15,0.55) {$\sigma$};

\draw[->,thick]
(-0.90,-0.32)
arc[start angle=-35,end angle=-315,x radius=0.32,y radius=0.52];

\node at (-0.42,0.18) {$\tau$};


\node at (1.95,0) {$\Longleftrightarrow$};


\fill[gray!25]
(4,-2.2) -- (5.7,-1.1) -- (5.7,2.2) -- (4,1.1) -- cycle;

\draw
(4,-2.2) -- (5.7,-1.1)
(5.7,-1.1) -- (5.7,2.2)
(5.7,2.2) -- (4,1.1)
(4,1.1) -- (4,-2.2);

\fill[gray!20]
(8,-2.2) -- (9.7,-1.1) -- (9.7,2.2) -- (8,1.1) -- cycle;

\draw
(8,-2.2) -- (9.7,-1.1)
(9.7,-1.1) -- (9.7,2.2)
(9.7,2.2) -- (8,1.1)
(8,1.1) -- (8,-2.2);

\shade[left color=blue!20,right color=blue!55,opacity=0.45]
(5.1,-0.7) rectangle (8.6,0.7);

\shade[inner color=blue!10,outer color=blue!50,opacity=0.6]
(5.1,0) ellipse (0.38 and 0.72);

\draw (5.1,0) ellipse (0.38 and 0.72);


\draw[thick,smooth cycle]
plot coordinates {
(6.60,0.72)
(6.35,0.58)
(6.20,0.20)
(6.28,-0.18)
(6.48,-0.55)
(6.82,-0.70)
(7.00,-0.30)
(6.95,0.18)
(6.82,0.55)
};

\fill[gray!10] (8.6,0) ellipse (0.38 and 0.72);
\draw (8.6,0) ellipse (0.38 and 0.72);

\draw[->,thick] (6.0,-1.28) -- (7.6,-1.28);

\node at (6.6,-1.55) {$\tau$};

\node at (6.5,1.02) {$\sigma$};

\end{tikzpicture}

\caption{Illustration of world-sheet duality relating the cylinder amplitude in the open sector and closed sector.}
\label{fig:worldsheet_duality}

\end{figure}

For the free bosonic string, the open-string boundary conditions are,

\begin{equation}
\text{Neumann boundary condition:} \qquad \partial_\sigma X^\mu \Big|_{\sigma=0,\pi}=0,
\end{equation}
which allows the string endpoint to move freely.
\begin{equation}
\text{Dirichlet boundary condition:} \qquad \delta X^\mu\Big|_{\sigma=0,\pi}=0,
\end{equation}
which fixes the endpoint position. Equivalently, since the endpoint is fixed in time,
\begin{equation}
\partial_\tau X^\mu\Big|_{\sigma=0,\pi}=0.
\end{equation}
Under the exchange
\begin{equation}
\sigma \leftrightarrow \tau,
\end{equation}
these become boundary conditions in the closed-string channel at fixed $\tau=0$
\begin{equation}
\partial_\tau X^\mu_{\text{closed}}\Big|_{\tau=0}|B_N\rangle =0,
\end{equation}
for Neumann conditions, and
\begin{equation}
\partial_\sigma X^\mu_{\text{closed}}\Big|_{\tau=0}|B_D\rangle =0,
\end{equation}
for Dirichlet conditions.\\

For a free closed bosonic string,
\begin{equation}
X^\mu(\tau,\sigma)
=
x^\mu
+
2\alpha' p^\mu \tau
+
i\sqrt{\frac{\alpha'}{2}}
\sum_{n\neq0}
\frac{1}{n}
\left(
\alpha_n^\mu e^{-in(\tau+\sigma)}
+
\tilde\alpha_n^\mu e^{-in(\tau-\sigma)}
\right),
\end{equation}
where the oscillators satisfy
\begin{equation}
[\alpha_m^\mu,\alpha_n^\nu]
=
m\delta_{m+n,0}\eta^{\mu\nu},
\qquad
[\tilde\alpha_m^\mu,\tilde\alpha_n^\nu]
=
m\delta_{m+n,0}\eta^{\mu\nu}.
\end{equation}

Substituting the mode expansion into the boundary conditions and evaluating at $\tau=0$ yields the oscillator relations,
\begin{equation}
\text{Neumann gluing condition:}\qquad(\alpha_n^\mu+\tilde\alpha_{-n}^\mu)|B_N\rangle=0,
\end{equation}
\begin{equation}
\text{Dirichlet gluing condition:}\qquad(\alpha_n^\mu-\tilde\alpha_{-n}^\mu)|B_D\rangle=0.
\end{equation}

These are called gluing conditions because they relate the two oscillator sectors at the boundary.

The corresponding boundary states are coherent squeezed states
\begin{equation}
|B_N\rangle
=
\mathcal N_N
\exp\left[
-\sum_{n=1}^\infty
\frac{1}{n}
\alpha_{-n}\cdot\tilde\alpha_{-n}
\right]
|0\rangle,
\end{equation}
and
\begin{equation}
|B_D\rangle
=
\mathcal N_D
\exp\left[
+\sum_{n=1}^\infty
\frac{1}{n}
\alpha_{-n}\cdot\tilde\alpha_{-n}
\right]
|0\rangle.
\end{equation}

These states encode the coupling of closed strings to D-branes. In particular, Neumann conditions correspond to freely propagating directions along the brane worldvolume, while Dirichlet conditions correspond to transverse directions fixed at the location of the brane.

In the tensionless limit, the worldsheet theory no longer possesses the conventional left-right decomposition of a two-dimensional conformal field theory. Nevertheless, an analogue of the boundary state construction survives. The usual oscillator gluing conditions are replaced by relations among the Carrollian oscillators $C_n^\mu$ and $\widetilde C_n^\mu$, leading to the induced vacuum conditions discussed in the Section \ref{sec:boundary_conditions}.

\section{Symplectic form, Poisson brackets and the Moyal $\star$-Product}\label{appendixA}
In this appendix, we review some basic ideas about symplectic form and Poission brackets \cite{Arnold:1989who,Nair:2024wyq}.
The symplectic formulation of classical mechanics provides a coordinate-independent
and geometrically natural language for describing dynamical systems.
Rather than treating coordinates and momenta separately, one regards the
\emph{phase space} as a smooth, even-dimensional manifold \( M \),
equipped with a closed and non-degenerate two-form \( \Omega \),
called the \emph{symplectic form}.
This two-form is the central object of Hamiltonian dynamics, from which
both the equations of motion and the structure of Poisson brackets emerge.

\subsection*{Symplectic form and its properties}

A \emph{symplectic manifold} \( (M, \Omega) \) is a differentiable manifold \( M \)
equipped with a differential two-form
\begin{equation}
\Omega = \frac{1}{2}\,\Omega_{\mu\nu}(q)\,dq^\mu \wedge dq^\nu,
\end{equation}
satisfying
\begin{equation}
d\Omega = 0, \qquad \det(\Omega_{\mu\nu}) \neq 0.
\end{equation}
The first condition expresses the closure of the form under exterior differentiation,
while the second ensures its non-degeneracy.
The non-degeneracy implies the existence of an inverse matrix \( \Omega^{\mu\nu} \),
defined by
\begin{equation}
\Omega^{\mu\alpha}\,\Omega_{\alpha\nu} = \delta^\mu_{\;\nu}.
\end{equation}

The closure condition \( d\Omega = 0 \) means that, at least locally,
there exists a one-form \( A \) such that
\begin{equation}
\Omega = dA, \qquad A = A_\mu(q)\,dq^\mu,
\end{equation}
called the \emph{symplectic potential} or \emph{canonical one-form}.
There is a gauge-like freedom
\begin{equation}
A \;\longrightarrow\; A + d\Lambda,
\end{equation}
for any smooth function \( \Lambda(q) \), which leaves \( \Omega \) invariant.

\subsection*{Canonical transformations and Hamiltonian vector fields}

Let, \( \xi = \xi^\mu(q)\,\partial_\mu \) be a vector field on \( M \).
The Lie derivative gives its action on the symplectic form,
\begin{equation}
\mathcal{L}_\xi \Omega = i_\xi d\Omega + d(i_\xi \Omega).
\end{equation}
A transformation generated by \( \xi \) is \emph{canonical}
if it preserves \( \Omega \), that is, \( \mathcal{L}_\xi \Omega = 0 \).
Since \( d\Omega = 0 \), this reduces to
\begin{equation}
d(i_\xi \Omega) = 0.
\end{equation}
Therefore, \( i_\xi \Omega \) is a closed one-form.
If the first cohomology group, \( H^1(M) \) is trivial, one can write
\begin{equation}
i_\xi \Omega = -df,
\end{equation}
for some function \( f(q) \) on \( M \).
This function \( f \) is called the \emph{generator}
of the canonical transformation, and \( \xi \) is the corresponding
\emph{Hamiltonian vector field}.
In local coordinates, the components of \( \xi \) can be written as
\begin{equation}
\xi^\mu = \Omega^{\mu\nu}\,\partial_\nu f.
\end{equation}
Hence, the flow generated by \( f \) acts on any other function \( g \)
through
\begin{equation}
\xi(g) = \Omega^{\mu\nu}\,\partial_\mu f\,\partial_\nu g.
\end{equation}
This operation will define the Poisson bracket.

Every smooth function on phase space generates a Hamiltonian flow
and thus corresponds to an infinitesimal canonical transformation. Observables are no longer passive quantities;
each defines a transformation preserving the symplectic structure.
In particular, the Hamiltonian function governing the energy
of the system generates the time evolution.

\subsection*{Poisson brackets}

Given two smooth functions \( f(q) \) and \( g(q) \) on \( M \),
their \emph{Poisson bracket} is defined by
\begin{equation}
\{f, g\} = \Omega^{\mu\nu}\,\partial_\mu f\,\partial_\nu g.
\end{equation}
This operation measures the infinitesimal change of one observable under the
canonical flow generated by the other.
For any function \( F(q) \), its variation under the flow of \( f \) is
\begin{equation}
\delta F = \{F, f\}.
\end{equation}

From the antisymmetry of \( \Omega^{\mu\nu} \), one has
\begin{equation}
\{f, g\} = -\{g, f\},
\end{equation}
and from the closure of \( \Omega \),
the Poisson bracket satisfies the \emph{Jacobi identity},
\begin{equation}
\{f, \{g, h\}\} + \{g, \{h, f\}\} + \{h, \{f, g\}\} = 0.
\end{equation}
The set of smooth functions on \( M \) thus forms a Lie algebra
under the Poisson bracket.

In local canonical coordinates \( (x^i, p_i) \),
where
\begin{equation}
\Omega = dp_i \wedge dx^i,
\end{equation}
one finds the familiar relations
\begin{equation}
\{x^i, x^j\} = 0, \qquad \{x^i, p_j\} = \delta^i_j, \qquad \{p_i, p_j\} = 0.
\end{equation}

\noindent\textbf{Remarks.}
It is often instructive to compare the symplectic formulation with other approaches.
If we apply the definition of the Poisson bracket directly to the phase-space
coordinates themselves, we find
\begin{equation}
\{q^\mu, q^\nu\} = \Omega^{\mu\nu}.
\end{equation}
This simple relation expresses that the ``basic Poisson brackets'' of the coordinates
are nothing but the inverse of the symplectic structure.
In other words, \( \Omega^{\mu\nu} \) plays the same role in the algebra of functions
that \( \Omega_{\mu\nu} \) plays in the geometry of the manifold.
This identification provides a clear link between the geometric and algebraic
descriptions of classical mechanics.

\subsection*{Symplectic potential and gauge analogy}

The canonical one-form \( A = p_i\,dx^i \) satisfies \( \Omega = dA \).
Under an infinitesimal canonical transformation generated by \( f \),
one finds
\begin{equation}
\delta_\xi A = \mathcal{L}_\xi A
= i_\xi dA + d(i_\xi A)
= -df + d(i_\xi A),
\end{equation}
so that, up to an exact differential,
\begin{equation}
A \rightarrow A + d\Lambda, \qquad \Lambda = i_\xi A - f.
\end{equation}
This transformation law is identical to that of a \( U(1) \) gauge potential,
with \( A \) playing the role of the connection and \( \Omega = dA \)
its curvature.
This analogy underlies the geometric interpretation of quantization,
where the symplectic form is identified with the curvature of a line bundle.

\subsection*{Darboux’s theorem and canonical coordinates}

A fundamental result of symplectic geometry is \emph{Darboux’s theorem},
which states that in the neighbourhood of any point on \( M \),
one can always find local coordinates \( (x^i, p_i) \)
such that
\begin{equation}
\Omega = dp_i \wedge dx^i.
\end{equation}
This means that all symplectic manifolds are locally equivalent;
they possess no local invariants.
The physical content of a Hamiltonian system is therefore encoded
not in the local form of \( \Omega \),
but in its global and topological properties.

The proof proceeds by induction.
Starting from a nonconstant function \( p_1(q) \),
one defines its conjugate \( x^1 \) via the flow generated by \( p_1 \),
so that \( \{x^1, p_1\} = 1 \).
Restricting \( \Omega \) to the submanifold \( p_1 = x^1 = 0 \)
yields a reduced two-form \( \Omega^* \)
on a \( (2n - 2) \)-dimensional subspace.
Repeating this process iteratively, one obtains all canonical pairs,
thereby reducing \( \Omega \) to the canonical Darboux form.

\subsection*{Poisson geometry and quantization}

The symplectic and Poisson structures together define the geometric
framework of classical mechanics.
A Poisson manifold generalizes this setting by allowing for degenerate brackets,
which naturally appear in systems with constraints or gauge symmetries.
The physical phase space is then obtained by quotienting out the degenerate directions,
leading to the Dirac bracket formalism.

At the quantum level, the Poisson algebra becomes the algebra of commutators:
\begin{equation}
\{f, g\} \;\longrightarrow\; \frac{1}{i\hbar}\,[\hat{f}, \hat{g}].
\end{equation}
This correspondence, the cornerstone of canonical quantization,
shows how the symplectic and Poisson structures provide the classical seeds of quantum mechanics.

\subsection*{The Moyal $\star$-Product}
The Moyal $\star$-product is a cornerstone of deformation quantization, providing a systematic way to replace the commutative algebra of classical observables with a noncommutative algebra. We closely follow \cite{Kontsevich:1997vb, Cornalba:2001sm, Herbst:2001ai} for the Moyal $\star$-Product. 
\subsection*{Definition}

On $\mathbb{R}^n$ equipped with a constant antisymmetric matrix $\Theta^{ij}$, the Moyal (or Groenewold–Moyal) $\star$-product between two smooth functions $f,g$ is defined by
\begin{equation}
(f\star g)(x)
= f(x)\exp\!\left(\frac{i}{2}\overleftarrow{\partial_i}\Theta^{ij}\overrightarrow{\partial_j}\right)g(x).
\end{equation}
Expanding the exponential gives
\begin{equation}
(f\star g)(x)
= \sum_{m=0}^{\infty}\frac{1}{m!}\left(\frac{i}{2}\right)^m
\Theta^{i_1 j_1}\cdots\Theta^{i_m j_m}
(\partial_{i_1}\cdots\partial_{i_m}f)(x)
(\partial_{j_1}\cdots\partial_{j_m}g)(x).
\end{equation}

\subsection*{Basic properties}

\paragraph{(1) Associativity.}
For constant $\Theta^{ij}$, the bidifferential operator
$\exp\big(\tfrac{i}{2}\overleftarrow{\partial_i}\Theta^{ij}\overrightarrow{\partial_j}\big)$
acts associatively, hence
\[
(f\star g)\star h = f\star(g\star h).
\]
This follows directly from the Baker–Campbell–Hausdorff formula for constant Poisson structures.

\paragraph{(2) Deformation of the commutative product.}
Expanding to second order in $\Theta^{ij}$:
\begin{equation}
f\star g = fg + \frac{i}{2}\Theta^{ij}\partial_i f\,\partial_j g
-\frac{1}{8}\Theta^{i_1 j_1}\Theta^{i_2 j_2}
\partial_{i_1}\partial_{i_2}f\,
\partial_{j_1}\partial_{j_2}g + \mathcal{O}(\Theta^3).
\end{equation}
Hence the $\star$-commutator reproduces the Poisson bracket at leading order,
\begin{equation}
[f,g]_\star \equiv f\star g - g\star f
= i\,\Theta^{ij}\partial_i f\,\partial_j g + \mathcal{O}(\Theta^3).
\end{equation}

\paragraph{(3) Trace property.}
For sufficiently decaying functions or periodic boundary conditions,
\begin{equation}
\int d^n x\, (f\star g)
= \int d^n x\, f g
= \int d^n x\, (g\star f).
\end{equation}
The equality holds because the extra terms in the expansion are total derivatives.

\paragraph{(a) Coordinate products.}
For coordinate functions $x^i$ one finds
\begin{equation}
x^i\star x^j = x^i x^j + \frac{i}{2}\Theta^{ij},
\qquad
[x^i,x^j]_\star = i\Theta^{ij}.
\end{equation}
Thus, the $\star$-product realizes the noncommutative coordinate algebra
\([x^i,x^j]=i\Theta^{ij}\).

\paragraph{(b) Plane waves.}
For $f(x)=e^{ik\cdot x}$ and $g(x)=e^{ip\cdot x}$,
\begin{equation}
e^{ik\cdot x}\star e^{ip\cdot x}
= e^{i(k+p)\cdot x}\,
e^{-\frac{i}{2}k_i\Theta^{ij}p_j}.
\end{equation}
Hence the $\star$-commutator of plane waves is
\begin{equation}
[e^{ik\cdot x},e^{ip\cdot x}]_\star
= 2i\,e^{i(k+p)\cdot x}\,
\sin\!\Big(\frac{1}{2}k_i\Theta^{ij}p_j\Big).
\end{equation}

\paragraph{(c) Gaussian example.}
Let $f(x)=e^{-\frac{1}{2}x^T A x}$ and
$g(x)=e^{-\frac{1}{2}x^T B x}$, where $A,B$ are symmetric.
Using the Fourier representation and the plane-wave composition rule,
one obtains another Gaussian
\begin{equation}
f\star g(x)
=\frac{1}{\sqrt{\det\!\left(1+\frac{1}{4}\Theta A\Theta B\right)}}
\exp\!\left[-\tfrac{1}{2}x^T C x\right],
\end{equation}
where $C$ is an explicit symmetric matrix depending on $A,B,\Theta$.
This identity is useful in heat-kernel and propagator computations in noncommutative field theory.

\subsection*{Relation to Weyl quantization}

The Moyal $\star$-product is the image of operator multiplication under the Weyl map $W$
\begin{equation}
W(f)\,W(g) = W(f\star g).
\end{equation}
In this sense, $\star$-multiplication is the pullback of operator composition to the space of functions, and quantizes the classical Poisson algebra determined by $\Theta^{ij}$.

\subsection*{Connection to the $B$-Field and noncommutative geometry}

In open string theory, a constant background $B$-field induces a noncommutative structure on the D-brane worldvolume with 
\begin{equation}
\Theta^{ij}
= - (2\pi\alpha')^2
\Big[(g+2\pi\alpha' B)^{-1}
B
(g-2\pi\alpha' B)^{-1}\Big]^{ij}.
\end{equation}
In the large-$B$ limit, this reduces to
$\Theta^{ij}\simeq B^{ij}$.
Therefore, the Moyal $\star$-product with parameter $\Theta^{ij}$ describes the effective noncommutative geometry of open string endpoints in the presence of a constant $B$-field.

\subsection*{Endpoint commutator}

For boundary coordinates $X^i(\tau)$, regarded as functions on the D-brane worldvolume,
\begin{equation}
[X^i(\tau),X^j(\tau)]_\star
= X^i\star X^j - X^j\star X^i
= i\,\Theta^{ij}.
\end{equation}
Substituting $\Theta^{ij}$ from the Seiberg--Witten relation gives the full dependence on $(g,B,\alpha')$.
In the tensionless limit, one recovers $[X^i,X^j]_\star=i~B^{ij}$,
as discussed in the main text.

\subsection*{Remarks}

\begin{itemize}
    \item The $\star$-product provides an explicit deformation quantization of the Poisson manifold with structure tensor $\Theta^{ij}$.
    \item It preserves associativity and reduces to the usual product in the commutative limit $\Theta^{ij}\to 0$.
    \item In string theory, the noncommutative phase factors $e^{-\tfrac{i}{2}k_i\Theta^{ij}p_j}$ appear in open-string amplitudes and account for the mixing between the endpoints induced by the $B$-field background.
\end{itemize}

\section{Worldsheet derivation of the noncommutative parameter}\label{appendixB}
For completeness, let us discuss the  derivation of the noncommutative parameter
$\Theta^{ij}$ as given by Seiberg and Witten \cite{Seiberg:1999vs}.
Their argument is based on the worldsheet operator formulation of the open string
propagating on a D-brane in the presence of a constant metric $g_{ij}$
and an antisymmetric $B$-field.
The mixed boundary condition induced by $B_{ij}$
leads to a deformation of the equal-time commutator of the boundary coordinates,
which is identified with the spacetime noncommutative parameter.

\subsection*{Boundary condition and reflection relation}

Consider an open tensile string on the upper half plane (UHP), with the boundary at $\text{Im}(z) = 0$, where $z=\tau +i \sigma$, and $\bar{z}=\tau -i \sigma$.
We decompose the embedding coordinates $(x^i)$ into holomorphic and anti-holomorphic parts as following,
\begin{equation}
x^i(z,\bar z) = X^i(z) + \widetilde X^i(\bar z).
\end{equation}
The open string boundary condition at $\text{Im{(z)}} = 0$ reads \cite{Seiberg:1999vs},
\begin{equation}
g_{ij}(\partial - \bar\partial)x^j + 2\pi\alpha' B_{ij}(\partial + \bar\partial)x^j 
\Big|_{z=\bar z} = 0.
\end{equation}
This couples the tangential and normal derivatives of $x^i$, and hence the holomorphic
and antiholomorphic components of the field.

Evaluated on the boundary $z=\bar z=\tau$, the condition becomes
\begin{equation}
(g+2\pi\alpha' B)_{ij}\,\partial X^j(\tau)
= (g-2\pi\alpha' B)_{ij}\,\bar\partial \widetilde X^j(\tau).
\end{equation}
Introducing the matrix
\begin{equation}
M = (g+2\pi\alpha' B)^{-1}(g-2\pi\alpha' B),
\end{equation}
This relation implies that the antiholomorphic coordinate can be written as a reflected image
of the holomorphic one:
\begin{equation}
\widetilde X^i(\bar z) = (M^{-1})^i{}_{\,j}\,X^j(z^*),
\qquad z^*=\overline{z}.
\label{eq:reflection}
\end{equation}
This `reflection rule' will be used to express the full propagator on the UHP
in terms of holomorphic correlators only.

\subsection*{Bulk correlators}

In the bulk of the UHP, the free correlators are standard:
\begin{align}
\langle X^i(z)X^j(w)\rangle &= -\alpha' g^{ij}\ln(z-w), \\
\langle \widetilde X^i(\bar z)\widetilde X^j(\bar w)\rangle &= -\alpha' g^{ij}\ln(\bar z-\bar w).
\end{align}
Using equation \eqref{eq:reflection}, the full open-string propagator can be decomposed into four terms:
\begin{align}
\langle x^i(z,\bar z)\,x^j(z',\bar z')\rangle
&=\langle X^i(z)X^j(z')\rangle
+\langle X^i(z)\widetilde X^j(\bar z')\rangle \nonumber\\
&\quad+\langle \widetilde X^i(\bar z) X^j(z')\rangle
+\langle \widetilde X^i(\bar z)\widetilde X^j(\bar z')\rangle.
\end{align}
Each term is evaluated as follows.  
The first is the ordinary holomorphic correlator:
\begin{equation}
\langle X^i(z)X^j(z')\rangle = -\alpha' g^{ij}\ln(z-z').
\end{equation}
The mixed term $\langle X^i(z)\widetilde X^j(\bar z')\rangle$ gives
\begin{align}
\langle X^i(z)\widetilde X^j(\bar z')\rangle
&= (M^{-1})^j{}_{\,l}\,\langle X^i(z) X^l(z'^*)\rangle \nonumber\\
&= -\alpha'\,g^{i l}(M^{-1})^j{}_{\,l}\,\ln(z-\bar z').
\end{align}
Similarly,
\begin{align}
\langle \widetilde X^i(\bar z) X^j(z')\rangle
&= (M^{-1})^i{}_{\,k}\,\langle X^k(z^*) X^j(z')\rangle \nonumber\\
&= -\alpha'\,(M^{-1})^i{}_{\,k}\, g^{k j}\,\ln(\bar z-z'),
\end{align}
and
\begin{align}
\langle \widetilde X^i(\bar z)\widetilde X^j(\bar z')\rangle
&= (M^{-1})^i{}_{\,k}(M^{-1})^j{}_{\,l}\,\langle X^k(z^*) X^l(z'^*)\rangle \nonumber\\
&= -\alpha'\,(M^{-1})^i{}_{\,k}\,g^{kl}\,(M^{-1})^j{}_{\,l}\,\ln(\bar z-\bar z').
\end{align}
Combining all four contributions, one obtains
\begin{equation}
\begin{aligned}
\langle x^i(z,\bar z)\,x^j(z',\bar z')\rangle
&= -\alpha'\Big[\, g^{ij}\ln(z-z') 
+ g^{i l}(M^{-1})^j{}_{\,l}\ln(z-\bar z') \\
&\quad + (M^{-1})^i{}_{\,k}\, g^{k j}\,\ln(\bar z-z') 
+ (M^{-1})^i{}_{\,k}\,g^{kl}\,(M^{-1})^j{}_{\,l}\,\ln(\bar z-\bar z')\Big].
\end{aligned}
\label{eq:2ptcorrelator}
\end{equation}
We now define $R=M^{-1}=(g-2\pi\alpha' B)^{-1}(g+2\pi\alpha' B)$.
Equation \eqref{eq:2ptcorrelator} then takes the compact form
\begin{equation}
\begin{aligned}
\langle x^i x^j\rangle
= -\alpha'\Big[ g^{ij}\ln(z-z') 
+ g^{ik}R^j{}_{\,k}\ln(z-\bar z') 
+ R^i{}_{\,k}g^{kj}\ln(\bar z-z') 
+ R^i{}_{\,k}g^{kl}R^j{}_{\,l}\ln(\bar z-\bar z')\Big].
\end{aligned}
\end{equation}
\subsection*{Symmetric and antisymmetric parts}
Introduce the symmetric and antisymmetric combinations
\begin{align}
A^{ij} &= \tfrac12\big(g^{ik}R^j{}_{\,k}+R^i{}_{\,k}g^{kj}\big), \\
B^{ij} &= \tfrac12\big(g^{ik}R^j{}_{\,k}-R^i{}_{\,k}g^{kj}\big).
\end{align}
The propagator can then be rewritten in a more transparent form:
\begin{align}
\langle x^i(z)x^j(z')\rangle
&= -\alpha'\Big[ g^{ij}\ln|z-z'| - g^{ij}\ln|z-\bar z'| \nonumber\\
&\quad + G^{ij}\ln|z-\bar z'|^2
+ \tfrac{1}{2\pi\alpha'}\,\Theta^{ij}\ln\!\frac{z-\bar z'}{\bar z - z'} \Big] + D^{ij},
\end{align}
where
\begin{align}
G^{ij} &= A^{ij} = \tfrac12\big(g^{ik}R^j{}_{\,k}+R^i{}_{\,k}g^{kj}\big), \\
\theta^{ij} &= 2\pi\alpha'\cdot 2B^{ij} 
= 2\pi\alpha'\big(g^{ik}R^j{}_{\,k}-R^i{}_{\,k}g^{kj}\big).
\end{align}
Inverting the above expressions using
$R=(g-2\pi\alpha' B)^{-1}(g+2\pi\alpha' B)$, one finds
\begin{align}
G_{ij} &= g_{ij}-(2\pi\alpha')^2 (B g^{-1} B)_{ij}, \\
\Theta^{ij} &= 2\pi\alpha'\Big(\frac{1}{g+2\pi\alpha' B}\Big)^{ij}_{\!A},
\end{align}
where the subscript $A$ denotes the antisymmetric part.
Here $G_{ij}$ is the open-string metric, and $\Theta^{ij}$ the antisymmetric
noncommutative parameter.

\subsection*{Boundary correlator and equal-time commutator}

On the boundary ($z=\tau$ and $z'=\tau'$ are real),
\begin{equation}
\ln\frac{z-\bar z'}{\bar z-z'} = i\pi\,\epsilon(\tau-\tau').
\end{equation}
The boundary two-point function becomes
\begin{equation}
\langle x^i(\tau)x^j(\tau')\rangle
= -\alpha' G^{ij}\ln(\tau-\tau')^2 + \tfrac{i}{2}\Theta^{ij}\,\epsilon(\tau-\tau').
\end{equation}
The antisymmetric piece proportional to $\epsilon(\tau-\tau')$ ($\epsilon(\tau)$ is the function that is $1$ or $-1$ for positive
or negative $\tau$)
defines the equal-time commutator
\begin{equation}
[x^i(\tau),x^j(\tau)] = i\,\Theta^{ij},
\end{equation}
which expresses the noncommutativity of the open-string end points.

\section{Two-point function in tensionless strings }\label{appendixC}
In the standard tensile string theory, the Polyakov action with a non-degenerate worldsheet metric leads to the familiar logarithmic propagator on the worldsheet \cite{Polchinski:1998rq, Green:2012oqa, Blumenhagen:2013fgp, Blumenhagen:2014sba},
\begin{equation}
\langle X^i(z,\bar z)\, X^j(w,\bar w)\rangle 
= -\,\frac{\alpha'}{2}\, g^{ij}\, \ln |z - w|^2,
\end{equation}
reflecting the two-dimensional nature of the conformal worldsheet with independent left- and right-moving sectors.  
In contrast, in the \emph{tensionless limit} ($T\!\to\!0$ or $\alpha'\!\to\!\infty$), 
the worldsheet metric becomes degenerate (Carrollian), and the dynamics reduce effectively to a one-dimensional theory along the worldsheet `time' direction.  
As a result, the two-point function becomes linear in the worldsheet time and ultralocal in the spatial coordinate.  

\subsection*{Action and equation for the Green’s function}
The free tensionless string action in the gauge $V^\alpha = (1, 0)$ takes the ultralocal form,
\begin{equation}
S =  \frac{1}{2}\int d\tau\, d\sigma \, g_{ij} \, \partial_\tau x^i \, \partial_\tau x^j,
\label{C:action}
\end{equation}
where $g_{ij}$ is the target-space metric. 
As the worldsheet metric is degenerate, the action contains only $\tau$-derivatives; 
the coordinate $\sigma$ acts merely as a label along the string, with no coupling between different $\sigma$-slices.

The quadratic operator associated with \eqref{C:action} is
\begin{equation}
\mathcal{D}_{ij}(\tau,\sigma;\tau',\sigma') 
= -\, g_{ij}\,\partial_\tau^2 \, \delta(\tau - \tau')\, \delta(\sigma - \sigma').
\end{equation}
The corresponding Green’s function $\mathcal{G}^{ij}(\tau,\sigma;\tau',\sigma')$ satisfies,
\begin{equation}
\mathcal{D}_{ik}(\tau,\sigma;\tau',\sigma')\, 
\mathcal{G}^{kj}(\tau',\sigma';\tau'',\sigma'') 
= \delta_i^{\;j}\,\delta(\tau-\tau'')\,\delta(\sigma-\sigma'').
\end{equation}
Because $\mathcal{D}$ acts only on $\tau$, the solution factorizes as
\begin{equation}
\mathcal{G}^{ij}(\tau,\sigma;\tau',\sigma') 
= g^{ij} \, \mathcal{G}_\tau(\tau-\tau') \, \delta(\sigma-\sigma'),
\end{equation}
where $\mathcal{G}_\tau(\tau-\tau')$ satisfies,
\begin{equation}
-\,\partial_\tau^2 \mathcal{G}_\tau(\tau-\tau') = \delta(\tau-\tau').
\label{C:reduced}
\end{equation}

\subsection*{Solution and normalization}

Equation \eqref{C:reduced} admits the symmetric solution on $\mathbb{R}$,
\begin{equation}
\mathcal{G}_\tau(\tau-\tau') = -\,\frac{1}{2}\,|\tau - \tau'|,
\end{equation}
since $\partial_\tau^2 \!\left(\tfrac{1}{2}|\tau-\tau'|\right) = \delta(\tau-\tau')$.  
This corresponds to the unique Green’s function that is symmetric under $\tau \leftrightarrow \tau'$ and vanishes at infinity.  
Homogeneous linear terms in $\tau$ or $\tau'$ correspond to zero modes and are fixed by the chosen boundary conditions.

If $\sigma$ is periodic, $\sigma \in [0,2\pi]$, 
the delta function should be replaced by the periodic distribution
\begin{equation}
\delta_{2\pi}(\sigma-\sigma') 
= \frac{1}{2\pi}\sum_{n\in\mathbb{Z}} e^{i n (\sigma-\sigma')}.
\end{equation}

\subsection*{The two-point function}

Combining the above results, the free two-point function of the embedding coordinates in the tensionless theory is
\begin{equation}
\langle x^i(\tau,\sigma)\, x^j(\tau',\sigma') \rangle
= -\,\frac{1}{2}\, g^{ij} \, |\tau - \tau'| \, \delta(\sigma - \sigma').
\label{C:2pt}
\end{equation}

This propagator exhibits the expected ultralocality of the degenerate worldsheet theory: 
different $\sigma$-points do not interact, and the correlation is linear (rather than logarithmic) in the worldsheet time separation.

Hence, while the tensile string propagator is logarithmic due to its two-dimensional conformal structure, 
the tensionless (Carrollian) limit yields the linear and ultralocal propagator \eqref{C:2pt}, 
signalling the collapse of left- and right-moving modes into a single degenerate worldsheet dynamics.

\providecommand{\href}[2]{#2}\begingroup\raggedright\endgroup

\end{document}